\documentclass[12pt,onecolumn,draftclsnofoot]{IEEEtran}

\usepackage{hyperref}
\usepackage{amsmath}
\usepackage{amsfonts}
\usepackage{soul}
\usepackage{amsmath}

\usepackage{cite}
\usepackage{amsmath,amssymb,amsfonts}
\usepackage[noend]{algpseudocode}
\usepackage{amsfonts}
\usepackage{graphicx}
\usepackage{textcomp}
\usepackage{gensymb}
\usepackage{xcolor}
\usepackage{tabularx}
\usepackage{color}
\usepackage{xfrac}
\usepackage{epstopdf}
\AppendGraphicsExtensions{.tif}
\usepackage{booktabs}
\usepackage{paralist}
\usepackage{enumitem}
\usepackage[export]{adjustbox}
\usepackage{multicol}
\usepackage{gensymb}
\usepackage{mathtools}
\usepackage{subcaption}
\usepackage{wrapfig}
\usepackage{circuitikz}
\usepackage{makecell}
\usepackage[normalem]{ulem}
\usepackage{soul}
\usepackage{comment}
\usepackage{diagbox}
\usepackage{fancyhdr}
\usepackage{algpseudocode}
\usepackage[linesnumbered,ruled,vlined]{algorithm2e}
\usepackage{multirow}
\usepackage{lipsum}
\DeclareUnicodeCharacter{2212}{-}
\def\BibTeX{{\rm B\kern-.05em{\sc i\kern-.025em b}\kern-.08em
    T\kern-.1667em\lower.7ex\hbox{E}\kern-.125emX}}

\usepackage{cite}

\ifCLASSINFOpdf
  
\else
  
\fi

\newtheorem{definition}{Definition}
\newtheorem{remark}{Remark}
\newtheorem{theorem}{Theorem}
\newtheorem{lemma}{Lemma}
\newtheorem{example}{Example}
\newtheorem{construction}{Construction}
\newtheorem{framework}{Framework}

\begin{document}

\title{Codes for Limited-Magnitude Probability Error in DNA Storage\\

\author{Wenkai Zhang,
        Zhiying Wang 
    }

\thanks{The authors are with the Center for Pervasive Communications and Computing (CPCC), University of California, Irvine, Irvine, CA 92697 USA (e-mail: wenkaiz1@uci.edu; zhiying@uci.edu).}
		}

\maketitle

\begin{abstract}
DNA, with remarkable properties of high density, durability, and replicability, 
is one of the most appealing storage media. Emerging DNA storage technologies use composite DNA letters, where information is represented by probability vectors, leading to higher information density and lower synthesizing costs than regular DNA letters. However, it faces the problem of inevitable noise and information corruption. This paper explores the channel of composite DNA letters in DNA-based storage systems and introduces block codes for {limited-magnitude probability errors} on probability vectors. First, outer and inner bounds for limited-magnitude probability error correction codes are provided. Moreover, code constructions are proposed where the number of errors is bounded by $t$, the error magnitudes are bounded by $l$, and the probability resolution is fixed as $k$. These constructions focus on leveraging the properties of limited-magnitude probability errors in DNA-based storage systems, leading to improved performance in terms of complexity and redundancy. In addition, the asymptotic optimality for one of the proposed constructions is established. Finally, systematic codes based on one of the proposed constructions are presented, which enable efficient information extraction for practical implementation.

\end{abstract}

\vspace{-2mm}
\section{Introduction}
In recent decades, DNA-based storage systems have been at the forefront of cutting-edge science and innovations \cite{1999dna, 2001dna, 2012dna, 2016dna, 2016dna1, DNA1}. These systems have gained significant appeal due to their high information density, durability, and replicability. 
The miniature size of DNA molecules allows for the storage of vast amounts of data in a compact form, making them an attractive option for data storage applications. The ability to store information with slow decay and degradation is another compelling advantage of DNA-based storage systems. Moreover, DNA strands can be efficiently replicated for information transfer, information restoration, and random information access.

The process of storing digital information on DNA involves several steps. First, the encoding process converts the digital data into a DNA sequence using the nucleotide alphabet (A, C, G, and T). 
Afterward, synthetic DNA molecules are created with the specified sequence representing the encoded information. 
The next step is to store the synthetic DNA molecules in suitable storage vessels, such as tubes or plates. 
When retrieval of the stored information is necessary, the sequencing process identifies the order of nucleotides in the DNA molecules. 
Lastly, decoding algorithms are applied to translate the DNA sequence back into the original digital information.
This comprehensive process of encoding, synthesizing, storing, sequencing, and decoding facilitates the use of DNA as a storage medium for digital information.

However, DNA-based storage suffers from high costs, especially during the synthesis process. In order to break through the theoretical limit of 2 bits per synthesis cycle for single-molecule DNA, the use of composite DNA letters was introduced by Anavy et al \cite{Anavy_DNA_letters, choi, Preuss, icc2022, yan2023, sabary2024survey}. 
A composite DNA letter is a representation of a position in a sequence that constitutes a mixture of all four standard DNA nucleotides with a pre-determined (scaled) \emph{probability vector} $(x_A,x_C,x_G,x_T)$, where $x_A,x_C,x_G,x_T$ are non-negative integers. Here, $k=x_A+x_C+x_G+x_T$ is fixed to be the \emph{resolution} parameter of the composite DNA letter. For example, a probability vector $(3, 3, 3, 3)$ represents a position in a composite DNA sequence of resolution $k = 12$. In this position, there is $\frac{1}{4}$ chance of seeing A, C, G or T. A composite DNA sequence is described by a \emph{probability word}. For example, the probability word $((3, 3, 3, 3),(6, 3, 3, 0),\dots)$ means that the first position in this sequence has equal probability of seeing each nucleotide, the second position has $\frac{1}{2}$, $\frac{1}{4}$, $\frac{1}{4}$ and $0$ chances for A, C, G and T, respectively, and so on.

Writing a composite DNA letter in a specific position of a composite DNA sequence involves producing multiple copies (oligonucleotides) of the sequence. Different DNA nucleotides are distributed across these synthesized copies based on the predetermined probabilities. It is important to note that the multiple copies of a fixed-length composite DNA sequence are produced simultaneously, resulting in a fixed synthesis cost but a higher information density as $k$ increases. Therefore, composite DNA offers a more affordable solution to DNA-based storage.

Reading a composite letter entails sequencing multiple copies and deducing the original probabilities from the observed frequencies\cite{Anavy_DNA_letters}.
The inference at any fixed position is affected by the sequencing depth $D$ (number of times the position is read), as well as synthesis and sequencing errors, resulting in the probability change $(x_A,x_C,x_G,x_T) \rightarrow (y_A,y_C,y_G,y_T)$. Specifically, let $\mathbf{p}=(p_A,p_C,p_G,p_T)$ be the probabilities (without the resolution constraint) after synthesis and sequencing errors, and let $\mathbf{p}'=(p'_A,p'_C,p'_G,p'_T)$ be the probabilities (with sequencing depth $D$) of the observed frequencies. Here, $\mathbf{p}'$ follows the multinomial distribution corresponding to the outcomes of a $4$-sided die rolled $D$ times, where the probability for each side is based on $\mathbf{p}$. Finally, $(y_A,y_C,y_G,y_T)$ equals the most probable $k$-resolution probabilities given $\mathbf{p}'$.

\begin{figure}
    \centering
    \includegraphics[width=0.3\linewidth]{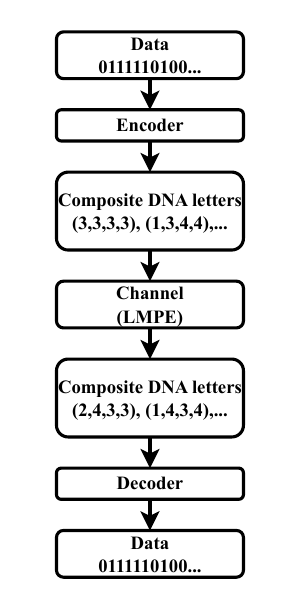}
    \caption{The illustration of composite DNA-based storage.}
    \label{The illustration of composite DNA storage.}
    \vspace{-0.6cm}
    
\end{figure}

The inferred probabilities $(y_A,y_C,y_G,y_T)$ and the original probabilities $(x_A,x_C,x_G,x_T)$ are usually close under correct operations and methods \cite{Anavy_DNA_letters}. 
For example, the original probabilities are $(3,3,3,3)$, then after synthesis and sequencing, the inferred probabilities are $(2,4,3,3)$. The probability ${x_A}$ is decreased by $1$ and the probability ${x_C}$ is increased by $1$, keeping the sum $k=12$. In this paper, errors are modeled as changing the probabilities of composite DNA letters in two directions (up or down) with limited magnitudes termed \emph{limited-magnitude probability errors (LMPE)}. 
The composite DNA-based storage is illustrated in Figure \ref{The illustration of composite DNA storage.}.  

We study block codes that correct limited-magnitude probability errors. A \emph{symbol} includes the probabilities of the four standard DNA nucleotides. The errors are parameterized by two integer parameters: 
$t$, the number of erroneous symbols in a codeword, and $l$, the maximum magnitude of errors in one direction in a symbol. More specifically, the magnitudes of the upward changes and downward changes are both within $l$. 

These errors are analogous to limited-magnitude errors over the set of integers modulo $q$. Asymmetric limited-magnitude errors with the binary alphabet $q=2$ are studied in \cite{Al_asymmetric/unidirect,M_Blaum,Bose_sys}. Asymmetric limited-magnitude error correction codes are proposed in \cite{Ahlswede_unidirectional} for the case of correcting all symbol errors within a codeword. In \cite{somecode, Four_Error, Cassuto_flash}, the authors study $q$-ary asymmetric limited-magnitude errors for multi-level flash. 
Perfect burst-correcting codes are provided for the limited-magnitude error motivated by the DNA-based storage system in \cite{wei}. 

Systematic coding schemes with minimum redundant symbols for correcting $q$-ary asymmetric limited-magnitude errors are shown in \cite{Elarief_0}.  To the best of our knowledge, the present work is the first to consider limited-magnitude probability errors.

In this paper, we first seek to bound the size of an error correction code for limited-magnitude probability errors using both the sphere-packing bound and the Gilbert-Varshamov bound. To apply the sphere-packing bound, one first defines the error ball for a given center word, which is the set of LMPE corrupted words from the center. 
Unlike the conventional alphabet of a finite field, the error balls for probability vectors are of different sizes because each probability must be between $0$ and $k$. Although this asymmetry exists, we find the smallest error-ball size is close to the largest error-ball size. Therefore, we use the smallest error-ball size to establish the upper bound of the size of the code. Moreover, to apply the Gilbert-Varshamov bound, for the conventional alphabet of a finite field, Hamming distance of at least $2t+1$ is used as a necessary and sufficient condition for a $t$-error correction code. 
For an LMPE correction code, however, we identify a sufficient distance condition and proceed to define and approximate the size of the corresponding ball of radius $2t+1$. For a large fixed $n$, we observe from the derived bounds that the code rates linearly decrease as the number of erroneous symbols 
$t$ increases, with $k$ and $l$ held constant. Additionally, the rates decrease logarithmically as the maximum magnitude of error $l$ increases, while $k$ and $t$ remain fixed.

Second, error correction codes are constructed for LMPE. Inspired by \cite{Cassuto_flash}, we note that given a stored probability vector, the observed probability vector has only a limited number of choices. Therefore, our constructions focus on protecting the ``class'' of the transmitted probability vector, which leads to a reduction in the redundancy and complexity. 
We present two ways to define the class of a probability vector, termed \emph{remainder class} and \emph{reduced class}, and introduce both \emph{normal} and \emph{improved} error correction codes for the class.
The resulting code is over a smaller finite field compared to naive error correction on the entire symbols, leading to lower computational complexity (see,  e.g.,  \cite{gashkov2013complexity} for discussion on the complexity for different finite field sizes). 

Third, we prove that the code with remainder classes and normal error correction asymptotically achieves the largest possible codes for LMPE, for large block length and large resolution. 
First, we establish upper and lower bounds on the size of our proposed code using remainder classes and an optimal normal error correction code.
Second, to obtain the converse, we relate an LMPE correction code over the probability vectors to an LMPE error correction code over the remainder classes, and the latter is further related to a normal error correction code over the remainder classes. In the end, we show the gap between the upper bound and lower bound of an LMPE correction code over the probability vectors that can be ignored for large block length and large resolution.

Finally, we present systematic LMPE correction codes. 

Our previous LMPE correction codes contain the information in the parity part. In order to separate the information from parity symbols, we define a Gray mapping between the probability vectors and the finite field elements. Specifically, for the information part, probability vectors are mapped to remainder classes (which are represented as finite field elements). For the parity part, probability vectors are mapped to finite field elements by the Gray code. We prove that limited-magnitude probability errors in probability vectors translate to a bounded number of finite field element errors by our Gray code, and error correction on the finite field elements is sufficient to recover the stored information. In the end, we also explore and analyze the redundancy comparison between systematic and non-systematic LMPE correction codes.

{\bf Organization.} We present the limited-magnitude probability error model of the composite DNA-based storage system in Section \ref{sec:statement}.
In Section \ref{sec:bound}, we provide the sphere-packing bound and the Gilbert-Varshamov bound for the family of limited-magnitude probability error correction codes.  Moreover, code constructions and the asymptotic optimality are shown in Section \ref{sec:cons}, followed by conclusions in Section \ref{sec:conclusion}.

{\bf Notation.} Vectors are denoted by bold letters. A sequence of vectors (or a word) is also denoted by a bold letter. $\binom{n}{k}$ denotes the binomial coefficient $n$ choose $k$. For a positive integer $i$, denote $[i]=\{1,2,\dots,i\}$. For two integers $i \le j$, denote $[i,j]=\{i,i+1,\dots,j\}$. For positive integers $i,j$, write $b = i \mod j$ as the remainder after $i$ is divided by $j$, where $0 \le b \le j-1$. The notation $b \equiv i \mod j$ means that integers $b$ and $i$ are congruent modulo $j$. The word ``probability'' is used to denote scaled probability, whose values are integers. Unless otherwise stated, $q$ is a prime power and $k \ge 6l$.

\section{Problem statement}\label{sec:statement}

We consider the problem of correcting limited-magnitude probability errors for probability words.
The word $\mathbf{x}=(\mathbf{x}_1,\mathbf{x}_2,\dots,\mathbf{x}_n)$ has $n$ symbols, and each \emph{symbol} or \emph{probability vector} $\mathbf{x}_i=(x_{i,1},x_{i,2},\dots,x_{i,m}), i \in [n],$ has $m$ (scaled) probability values satisfying
\begin{align}
&0 \le {x}_{i,j} \leq k, j\in[m], \label{eq:prob1}\\
&\sum_{j=1}^{m}{x}_{i,j}=k. \label{eq:prob2}
\end{align}
Here, $k$ is the fixed resolution parameter to keep the summation of probabilities to 1, namely, $\sum_{j=1}^{m}\frac{x_{i,j}}{k}=1$. 
The set of all symbols satisfying \eqref{eq:prob1} and \eqref{eq:prob2} is denoted by $\mathcal{X}$. One can easily check \footnote{The following formulas are used throughout the paper. Let $k, r$ be some positive integers.
The number of non-negative integer choices of $(x_1,x_2,\dots,x_r)$ to satisfy $x_1+x_2+\dots+x_r=k$ is $\binom{k+r-1}{r-1}$, and the number of positive integer choices of $(x_1,x_2,\dots,x_r)$ to satisfy $x_1+x_2+\dots+x_r=k$ is $\binom{k-1}{r-1}$.} 
that its size is $|\mathcal{X}|=\binom{k+m-1}{m-1}$. The set of words of length $n$ is $\mathcal{X}^n$.

Assume $\mathbf{x} \in \mathcal{X}^n$ is the transmitted codeword.
Denote by $\mathbf{y}=(\mathbf{y}_1,\mathbf{y}_2,\dots,\mathbf{y}_n) \in \mathcal{X}^n$ the received word. 
Denote by $\mathbf{e}=\mathbf{y}-\mathbf{x}=(\mathbf{e}_1,\mathbf{e}_2,\dots,\mathbf{e}_n)$ the error vector, and $e_{i,j}$ its probability difference values, $i \in [n], j \in [m]$. It is apparent that for all $i \in [n]$,
\begin{align*}
    \sum_{j=1}^{m} e_{i,j}=0.
\end{align*}
It should be noted that the set of \emph{valid errors} $\mathbf{e=y-x}$ varies as the codeword $\mathbf{x}$ or the received word $\mathbf{y}$ changes. In particular, 
\begin{align}
    & -x_{i,j} \le e_{i,j}\le k-x_{i,j} ,\label{eq:valid1}\\
    & y_{i,j}-k \le e_{i,j} \le y_{i,j}.\label{eq:valid2}
\end{align}
We point out that such asymmetry of the valid errors must be carefully handled when deriving the bounds on the limited-magnitude probability error correction codes in Section \ref{sec:bound}. 

\begin{definition}[Limited-magnitude probability error (LMPE)]\label{def:error}
A probability error $\mathbf{e}$ is called an $(l,t)$ LMPE if the number of symbol errors is at most $t$, and the error magnitude of each symbol is at most $l$. Formally,
\begin{align}
    &|\{i \in [n]: \mathbf{e}_i \neq \mathbf{0}\}| \le t,\\
    &\sum_{j \in [m]: e_{i,j}>0} e_{i,j} = \sum_{j \in [m]: e_{i,j}<0} |e_{i,j} | \le l, \forall  i \in [n]. \label{eq:5}
\end{align}

In situations where either the transmitted codeword $\mathbf{x} \in \mathcal{X}^n$ or the received word $\mathbf{y} \in \mathcal{X}^n$ is given, LMPE only refers to the valid errors satisfying \eqref{eq:valid1} or \eqref{eq:valid2} unless stated otherwise. 
\end{definition}

Equation \eqref{eq:5} indicates that the upward or downward error terms sum to at most $l$. Equivalently, it requires that the \emph{l}$_1$-norm of $\mathbf{e}_i$ is no more than $2l$, and the sum of the upward errors equals that of the downward errors.
In this paper, we consider the case of $m=4$, which fits the composite DNA-based storage application. 
For example, when $k=12$ and $n=3$, the transmitted word is  $\mathbf{x}$=$((3,3,3,3),(2,4,3,3),(1,7,2,2))$, 
the received word is  $\mathbf{y}$=$((1,3,\underline{4},\underline{4}),(1,\underline{5},3,3),(1,7,2,2))$. Here, the upward errors are underlined, the number of symbol errors is $t=2$, and the error magnitude is $l=2$.

Similar to \eqref{eq:5}, an \emph{$l$-limited-magnitude probability error} $\mathbf{e}=(e_1,e_2,e_3,e_4)$ for one symbol should satisfy
\begin{align}
    \sum_{j\in[4]:e_j > 0} e_j = ~~~\sum_{j\in[4]:e_j < 0} |e_j| \le l. \label{eq:e_upward}
\end{align}

\begin{lemma}\label{lem:limited error}
$\mathbf{e}=(e_1,e_2,e_3,e_4)$ is an $l$-limited-magnitude probability error for one symbol, if and only if 
\begin{align}
\sum_{j=1}^4 e_j= 0,\\
    \sum_{j=1}^4 |e_j| \le 2l. \label{eq:e_sum}
\end{align}
\end{lemma}
\begin{IEEEproof}
Given $\sum_{j=1}^4 e_j= 0$,
equivalently,
\begin{align}
    \sum_{j\in[4]:e_j > 0} e_j = \sum_{j\in[4]:e_j < 0} |e_j|.
\end{align}

If $\mathbf{e}$ is an $l$-limited-magnitude probability error, then adding the two sums in \eqref{eq:e_upward} gives \eqref{eq:e_sum}.

If $\mathbf{e}$ is an error with magnitude more than $l$, then each sum in \eqref{eq:e_upward} is more than $l$. Hence,
\begin{align*}
    l < \sum_{j\in[4]:e_j > 0} e_j =  \sum_{j\in[4]:e_j < 0} |e_j|. 
\end{align*}
Hence, $\sum_{j=1}^4 |e_j| > 2l$ as desired.
\end{IEEEproof}

\begin{definition}[LMPE correction code]
    An  \emph{$(l,t)$ LMPE correction code} $\mathcal{C} \subseteq \mathcal{X}^n$ is defined by an encoding function $Enc: \{0,1\}^K \to \mathcal{C}$ and a decoding function $Dec: \mathcal{X}^n \to \{0,1\}^K$. Here, $K$ is the information length in bits and $n$ is the codeword length in symbols. For any binary information vector $\mathbf{u}$ of length $K$, and any $(l,t)$ LMPE vector $\mathbf{e}$ of length $n$ such that $Enc(u) + e \in \mathcal{X}^n$, the functions should satisfy
\begin{align*}
    Dec(Enc(\mathbf{u})+\mathbf{e})= \mathbf{u}.
\end{align*}
\end{definition}

\section{Bounds on LMPE correction codes}\label{sec:bound}

In this section, we present the sphere-packing bound and the Gilbert-Varshamov bound on the code size for correcting $(l,t)$ LMPE. 

First, we define a distance that captures the capability of an LMPE correction code.

\begin{definition}[Geodesic distance]\label{def:distance}
Let $l$ be a fixed positive integer.
Consider a graph whose vertices are the set of words in $\mathcal{X}^n$. Two vertices $\mathbf{x},\mathbf{z}$ are connected by an edge if their difference $\mathbf{e}=\mathbf{z}-\mathbf{x}$ is an $(l,t=1)$ LMPE. The Geodesic distance $d_l({\mathbf{x},\mathbf{z}})$ of any two words $\mathbf{x},\mathbf{z}$ is defined as the length of the shortest path between them. 
\end{definition}

The geodesic distance is indeed a distance metric. 

\begin{remark}\label{remark:1}
Note that if a codeword $\mathbf{x}$ is transmitted and an $(l,t)$ LMPE occurs, then the received word $\mathbf{y}$ must satisfy $d_l(\mathbf{x},\mathbf{y})\le t$. On the other hand, if $d_l(\mathbf{x},\mathbf{y}) = t$, the error may not be an $(l,t)$ LMPE, because there can be less than $t$ symbol errors but some symbol errors can have the magnitude more than $l$.
\end{remark}

The distance gives a sufficient condition for the number of limited-magnitude symbol errors correctable by a code $\mathcal{C}$, for a fixed error magnitude $l$. 

\begin{theorem}\label{thm:distance}
A code $\mathcal{C} \subseteq \mathcal{X}^n$ can correct any $(l,t)$ LMPE if, for all distinct $\mathbf{x},\mathbf{z} \in \mathcal{C}$, $d_l(\mathbf{x},\mathbf{z}) \geq 2t+1 $.
\end{theorem}
\begin{IEEEproof}
Assume the code has a minimum distance of $2t+1$, but an $(l,t)$ LMPE is not correctable. Then there exist two codewords $\mathbf{x},\mathbf{z}$ and a received word $\mathbf{y}$ such that $d_l(\mathbf{x,y})\le t, d_l(\mathbf{z,y}) \le t$. Therefore, $d_l(\mathbf{x},\mathbf{z}) \le d_l(\mathbf{x},\mathbf{y})+ d_l(\mathbf{z,y}) \le 2t$, which is a contradiction to the minimum distance $2t+1$.
\end{IEEEproof}

Let $A_{prob, LMPE}(n,k,t,l) $ denote the maximum possible size of an $(l,t)$ LMPE correction code, whose codeword length is $n$ and resolution is $k$.

{\bf Sphere-packing bound.}
Define the error ball as the set of all possible erroneous received words for a codeword under $(l,t)$ LMPE. The sphere-packing bound is based on the following fact: the error balls for different codewords should be disjoint, and the union of all error balls is a subset of all the words.
Due to the constraints in \eqref{eq:valid1} \eqref{eq:valid2}, if a probability value is smaller than $l$ (or larger than $k-l$), the downward (or upward) error magnitude becomes smaller than $l$ as well, resulting in different error ball sizes. To accommodate the issue, we use the smallest error ball size for all the codewords.

\begin{theorem}\label{thm:sphere_packing}
    For $k\ge 4l$, the sphere-packing upper bound on the size of the LMPE correction code is  
    \begin{align}
    A_{prob, LMPE}(n,k,t,l)  \leq \frac{\binom{k+3}{3}^{n}}{ \binom{n}{t} \left(\frac{1}{6}\right)^{t} l^{3t}}\label{thm: eq: sephere}.    
\end{align}
\end{theorem}

\begin{IEEEproof}
    The detailed proof can be found in Appendix \ref{A}.
\end{IEEEproof}

To obtain the expression in \eqref{thm: eq: sephere}, low-order terms with respect to $n,l$ are removed in the denominator, leading to a relaxed upper bound (See Appendix \ref{A} for details). When $n,l$ are small, one can use the exact expression \eqref{eq: 66} in Appendix \ref{A}.

{\bf Gilbert-Varshamov bound.}
By Theorem \ref{thm:distance}, the distance being at least $2t+1$ is a sufficient condition to correct $(l,t)$ LMPE. Therefore, the Gilbert-Varshamov bound implies that the code size is lower bounded by the size of $\mathcal{X}^n$ divided by the size of a ball of radius $2t$, where the ball is defined as the set of words whose geodesic distance is at most $2t$ from the center. Note that this ball of radius $2t$ is larger than the error ball of $(l,2t)$ LMPE.

In the following theorem, we write $A \gtrsim B$ if $\lim_{l \to \infty}\frac{A}{B} \ge 1$. 
This notation is to simplify our final expression so that only the highest order terms with respect to $l$ are included. 

\begin{theorem}\label{thm:Gilbert}
    For $n > \frac{1}{l^3}(2t)^2 3.02^{2t}, l \gg 1$, 
    The Gilbert-Varshamov lower bound on the size of the LMPE correction code is
\begin{align}
A_{prob, LMPE}(n,k,t,l)  \gtrsim \frac{\binom{k+3}{3}^{n}}{(2t) \binom{n}{2t} \left(\frac{10}{3}\right)^{2t} l^{6t}}\label{thm: eq: GV}.   
\end{align}

\end{theorem}
\begin{IEEEproof}
    The detailed proof can be found in Appendix \ref{app:Gilbert}.
\end{IEEEproof}

\begin{remark}
One observation from the proof of Theorem \ref{thm:Gilbert} is that, for large $n$, the geodesic-distance ball of radius $d$ is almost the same as the error ball with $(l,d)$ LMPE, despite their difference as in Remark \ref{remark:1}. 
Therefore, even though geodesic distance provides only a sufficient condition for the existence of LMPE correction code, our Gilbert-Varshamov bound is asymptotically as tight as a bound using a necessary and sufficient condition. 
In particular, if $n > \frac{1}{l^3}d^2 3.02^d$, almost all error patterns in the geodesic-distance ball have the errors spread across $d$ distinct symbols, which corresponds to the LMPE ball. 
The above claim can be verified by observing that for $n > \frac{1}{l^3}d^2 3.02^d$, $V(d)$ is the dominant term in the geodesic-distance ball (see Appendix \ref{app:Gilbert} for the definition of $V(d)$ and Equation \eqref{eq: ratio} to see $V(d)$ is dominant), and all other terms can be neglected. 
\end{remark}

\begin{figure}[h]
    \centering
    \includegraphics[scale=0.7]{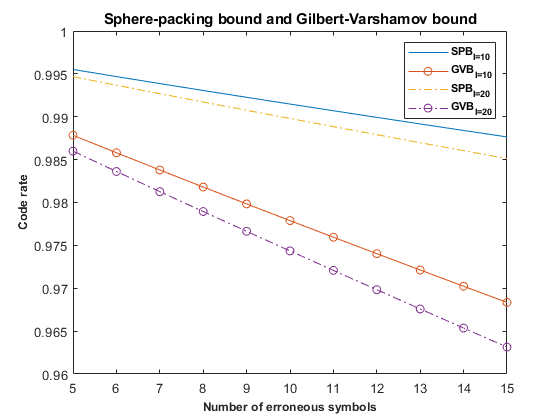}
    \caption{Bounds on LMPE correction code with $n= 1023$, $k=100$, $l=10$ or $20$. The horizontal axis is the number of erroneous symbols $t$, and the vertical axis is the code rate. SPB and GVB represent sphere-packing (upper) bound and Gilbert-Varshamov (lower) bound, respectively.}
    \label{fig:bound}
\end{figure}

Using \eqref{thm: eq: sephere} and \eqref{thm: eq: GV}, the sphere-packing bound and the Gilbert-Varshamov bound with different parameters are shown in Figure \ref{fig:bound}. For large $n$, when the number of erroneous symbols $t$ increases, the code rates of these two bounds linearly decrease. Moreover, given fixed $k,t$ and large $n$, the code rate decreases logarithmically with $l$. Numerically, when $n=1023, t=15$, the difference between the sphere-packing bound and the Gilbert-Varshamov bound is $1.95\%$ and $2.23\%$ for $l=10,l=20$, respectively.

\vspace{-0.2cm}

\section{Constructions}\label{sec:cons}
While the Gilbert-Varshamov bound is based on a non-explicit construction, we seek to provide explicit code constructions in this section. 
First, we use an example to illustrate the coding problem and introduce the main idea of the code constructions. Then we give the framework of the proposed codes for the limited-magnitude probability errors. Based on the framework, we provide three code constructions: remainder class codes, reduced class codes, and codes based on the improved Hamming code. 
The main concept behind our constructions is to categorize probability vectors into different ``classes''. This categorization allows us to protect the transmitted probability vectors effectively and enhance the code rate. 
Afterwards, we prove that our remainder class code construction is asymptotically optimal. 
In the end, in order to make the remainder class codes more practical, we present a construction of systematic remainder class codes.  

\subsection{Framework and example}
In this section, we present a general framework for our constructions and provide an example to illustrate it. All of our proposed constructions adhere to this framework. Before providing the framework, we define some necessary terms.

\begin{definition}[Quotient vector and remainder vector]\label{def: quo&rem}
We define the \emph{quotient vector} $(a_1,a_2,a_3,a_4)$ and the \emph{remainder vector} $(b_1,b_2,b_3,b_4)$ from the probability vector $(x_1,x_2,x_3,x_4)$ through division by $2l+1$, where $0 \le x_i \le k, \sum_{i=1}^{4}x_i=k$. Each probability value $x_i$ is divided by $2l+1$, and can be represented by the quotient $a_i$ and the remainder $b_i$, $0 \le b_i < 2l+1$, for $i \in [4]$: 
 \begin{align*}
      x_i = (2l+1)a_i + b_i, ~~ i \in[4].
  \end{align*}
\end{definition}

In our several constructions, we use  remainder vector to define the ``class'' of a symbol. ''classes'' refer to the distinct remainder vectors, which are used to differentiate between the remainder vectors. "class or classes index" refers to one distinct remainder vector or one distinct set of remainder vectors.

\begin{remark}\label{re: classes}
As defined in Definition \ref{def: quo&rem}, each remainder $b_i = x_i \mod (2l+1), i \in [4],$ can take on $2l+1$ possible values. Additionally, for every value of $b_1,b_2$ and $b_3$, the value of $b_4$ is uniquely determined by the equation $\sum_i b_i = k \mod (2l+1)$.
Therefore, there are at most $(2l+1)^3$ classes when $k \ge 6l$, since $b_1,b_2,b_3$ can be randomly chosen from $[2l]$. 
Denote the number of classes by $q_c$ for fixed $k$ and $l$. In our constructions, we will require a finite field size $q \le q_c$ for constructing codes to protect the classes. In the following, we only write for the case of $q_c=(2l+1)^3$. However, similar arguments easily generalize to any other $q_c$. 
\end{remark}
Now we are ready to present our framework.
\begin{framework}\label{cnstr:framework} The LMPE correction code framework includes one classification and two coding steps.

{\bf Symbol classification:} Symbols in $\mathcal{X}$ are mapped to ``classes'' such that any $l$-limited-magnitude symbol error changes the ``class'' of the symbol.

{\bf First layer:}
Construct a $t$-error correction code whose codeword symbols are the class indices. Thus, the decoder can identify the locations of the erroneous probability vectors.

{\bf Second layer:}
Construct a code such that given the correct classes and the received symbols, the original symbols can be recovered.

\end{framework}

Based on this framework, we can establish different coding schemes to protect the information in the limited-magnitude probability error channel. In the following, we provide our first example. This example makes use of Hamming codes over a finite field, which are perfect $1$-error correcting codes. For the sake of completeness, we briefly review $q$-ary Hamming code \cite{hamming,H} below.

The $(\frac{q^r-1}{q-1},\frac{q^r-1}{q-1}-r)$ Hamming code over $GF(q)$ has codeword length $\frac{q^r-1}{q-1}$, information length $\frac{q^r-1}{q-1}-r$ and $r$ redundant symbols. 
The Hamming code is defined by its parity check matrix of size $r \times \frac{q^r-1}{q-1}$, whose columns are all non-zero pair-wise independent columns of length $r$. 

\begin{figure}
    \centering
    \includegraphics[scale=0.6]{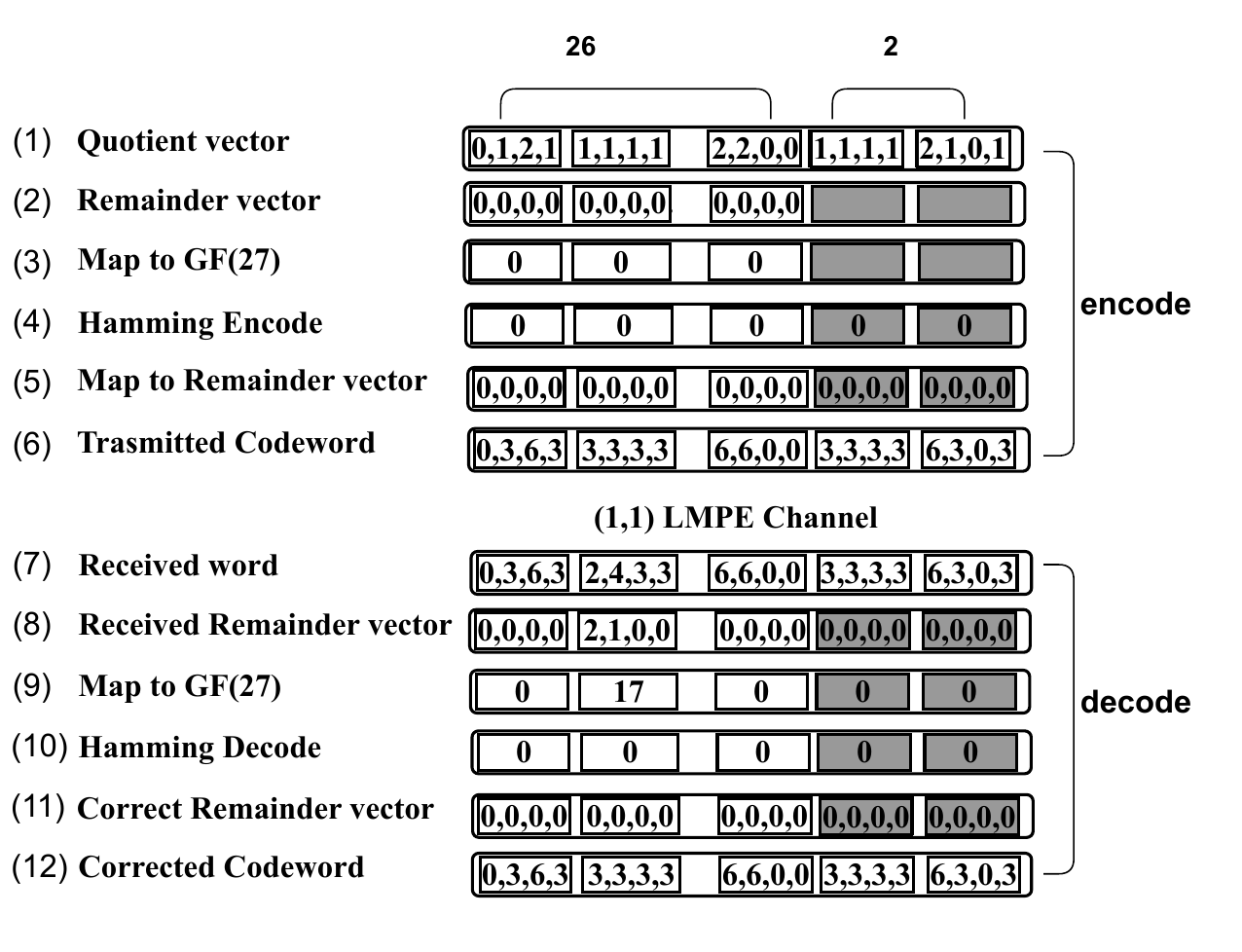}
    \caption{An $(l=1,t=1)$ LMPE correction code with $n=28$, resolution $k=12$, based on the $27$-ary $(28,26)$ Hamming code. This figure demonstrates the encoding, corruption, and decoding of only one example codeword. Assume $(0,0,0,0)$ is mapped to $0$ in $GF(27)$, and $(2,1,0,0)$ is mapped to $17$ in $GF(27)$. The gray color denotes the parities.
     In Rows (1) and (2), all $28$ quotient vectors and the first $26$ remainder vectors are mapped from the information messages.
     In Row (3), we map the first 26 remainder vectors to elements in $GF(27)$. In Row (4) we use the $27$-ary $(28,26)$ Hamming code to generate two $0$'s in $GF(27)$ as parities (last two symbols), which are mapped back to remainder vectors $(0,0,0,0)$ in Row (5). We combine the remainder vectors and the quotient vectors to form transmitted probability vectors in Row (6). 
     Row (7) denotes the corrupted word with a limited-magnitude probability error in the $2$nd symbol.
     In Row (8), we get the remainders through dividing by $3$. In Row (9), remainder vectors are mapped to elements in $GF(27)$. Then we use the $27$-ary $(28,26)$ Hamming code to decode in  Row (10), and the corresponding correct remainder vectors are in Row (11). Based on the received probability words in Row (7) and the correct remainder vectors in Row (11), we form the corrected codewords in Row (12). 
    }
    \label{fig:example1}
\end{figure}

\begin{example}\label{example:1}
We construct an $(l=1,t=1)$ LMPE correction code over probability vectors $\mathcal{X}$ for resolution $k=12$, illustrated in Figure \ref{fig:example1}. The code length is $n=28$.

When $k=12$, the quotient vector for a symbol is in the set $\{(a_1,a_2,a_3,a_4):\sum_{j=1}^{4} a_j \leq 4, a_j \in [0,4], j \in [4]\}$, and the remainder vector belongs to the set  $\{(b_1,b_2,b_3,b_4):\sum_{j=1}^{4} b_j \in\{ 0,3,6\}, b_j \in [0,2], j \in [4]\}$. Moreover, according to Remark \ref{re: classes}, we know the remainder vectors can have $(2l+1)^3=27$ ``classes''.

The code is described below.

{\bf Symbol classification:} 
Symbols in $\mathcal{X}$ are classified by their remainder vectors.

{\bf First layer:}
We arbitrarily map all the remainder vectors to elements of $GF(27)$ (e.g., see Table I). We use $27$-ary $(28,26)$ Hamming code to protect the remainder vectors. 

{\bf Second layer:}
Not needed.

Then we introduce the LMPE channel and encoding/decoding method of our example. 

{\bf Encoding:} The information is represented by $28$ quotient vectors and $26$ remainder vectors. The remainder vectors are mapped to $GF(27)$, and two additional elements in $GF(27)$ are generated from the $(28,26)$ Hamming code over $GF(27)$. Combining the $28$ pairs of quotient and remainder vectors, we obtain $28$ codeword symbols over probability vectors.

{\bf LMPE channel:} When $(l=1,t=1)$ limited-magnitude probability error occurs, at most one remainder vector is corrupted, and the upward and the downward error magnitude is 1. 

{\bf Decoding:} From the received symbols, compute the quotient vectors and the remainder vectors. Then the remainder vectors are mapped to elements in $GF(27)$. With the Hamming code over $GF(27)$, the $t=1$ error on the word over $GF(27)$ can be corrected, and correspondingly the corrupted remainder vector can be corrected.
Since the probability error has only magnitude $1$, it can be recovered solely by the correct remainder vectors and the received symbols (the quotient vectors do not need to be corrected). 

In Figure \ref{fig:example1}, the second transmitted symbol $(3,3,3,3)$ corresponds to the remainder vector $(0,0,0,0)$, and is changed to the received symbol  $(2,4,3,3)$ due to noise, whose remainder vector is $(2,1,0,0)$, or $17$ in $GF(27)$. The corrected remainder vector $(0,0,0,0)$ can be found due to the Hamming decoder. Given the received symbol $(2,4,3,3)$ and the correct remainder vector $(0,0,0,0)$, it can be seen that the only possible transmitted symbol with error magnitude $l=1$ is $(3,3,3,3)$.

We note that the quotient vectors and the remainder vectors are not independent. When the remainder vector satisfies $\sum_{j=1}^4b_j=6$, we must have $\sum_{j=1}^4 a_j=2$ because $k=12$. In this case, we have the least number of possible quotient vectors $(a_1, a_2, a_3, a_4)$, which is $\binom{2+4-1}{4-1}=10$.
Given the first $26$ information symbols, the last $2$ parity remainder vectors will be fixed. In order to accommodate the worst-case scenario, we only allow $10^2$ possible information messages represented by the last two quotient vectors. For example, in Figure \ref{fig:example1}, the last two quotient vectors $(1,1,1,1),(2,1,0,1)$ correspond to one of these $10^2$ messages. 
The code rate is $(26 \log_2 \binom{k+3}{3} + \log_2 10^2)/(28\log_2 \binom{k+3}{3}) = 0.955$ for $k=12$. 
According to Theorem \ref{thm:sphere_packing}, the sphere-packing upper bound on the code rate is $0.991$. Based on Theorem 3, Gilbert–Varshamov lower bound on the rate is $0.921$.

\end{example}

\begin{table}
\centering
\caption{The mapping between remainder vectors and $GF(27)$. Each element of $GF(27)$ is denoted by an integer between $0$ and $26$.}
\label{27-ary}

\begin{tabular}{|cc|cc|cc|} 
 \hline

 Remainder vector & Element & Remainder vector&Element& Remainder vector&Element \\
 \hline
 0000 & 0& 1110& 1& 2220 &2  \\ 
 \hline
0111&3 & 1221 & 4& 2001& 5 \\ 
 \hline
0222& 6 &1002&7 & 2112& 8\\
 \hline
0012& 9& 1122& 10& 2202&11 \\
 \hline
 0021& 12& 1101& 13& 2211&14 \\
  \hline
 0210& 15& 1020& 16& 2100&17\\ 
 \hline
0102& 18& 1212& 19& 2022&20 \\
 \hline
0201& 21& 1011& 22& 2121&23 \\
  \hline
 0120& 24& 1200& 25& 2010&26\\
 \hline

\end{tabular}
\end{table}

Our example is designed to illustrate our framework and our main idea of encoding and decoding. In the following, we are ready to provide more general two-layer codes based on our framework to correct limited-magnitude probability errors.

\subsection{Remainder class codes}

Our method of correcting errors only on the remainders is inspired by \cite{Cassuto_flash}, where the modulo operation was introduced for limited-magnitude errors on integers modulo $q$. 

One of the differences in our construction is that it requires the additional steps of mapping between remainder vectors and finite field elements, in order to accommodate for the alphabet of probability vectors.

We can select any $t$-error correction code over the class alphabet for the first layer. For illustration purposes, we use the well-known BCH codes \cite{BCH1,BCH2,BCH}.
BCH codes include Hamming codes as a special case. For a given field $GF(q)$ and an integer $w$,  $q$-ary BCH code has codeword length $n=q^w -1$, minimum distance at least $2t+1$, and at most $2wt$ parity check symbols.

The next $(l,t)$ LMPE correct code is a generalization of Example \ref{example:1} using BCH codes to protect remainder classes. Its correctness can be seen similarly as Example \ref{example:1}. 

\begin{construction}[Remainder classes]\label{cnstr:remainder classes} \\
{\bf Symbol classification:} Divide each probability vector by $(2l+1)$ to get the remainder vector based on Definition \ref{def: quo&rem}, and accordingly obtain $(2l+1)^3$ classes. \\
{\bf First layer:} Apply a BCH code over $GF(q)$ to the remainder vectors, whose length is $n= q^w -1$, $q$ is a prime power, $q \leq(2l+1)^3$, and the code distance is $2t+1$. \\
{\bf Second layer:} Not needed.
\end{construction}

\begin{remark}\label{remark: 4}
Unless otherwise stated, we assume that $(2l+1)^3$ is a prime power. If $(2l+1)^3$ is not  {a prime power}, our Construction \ref{cnstr:remainder classes} still works. 

For encoding,  {since $(2l+1)^3$ is not a prime power, we will choose $q < (2l+1)^3$.} The number of remainder classes is $(2l+1)^3$, we can still map finite field elements to a subset of remainder vectors. 

For decoding, we emphasize that our construction can easily handle an error where a received remainder vector cannot be mapped back to the finite field element, because this error can be treated as an erasure. 
Therefore, if at most $t$ limited-magnitude errors occur, we get $t'$ errors and at most $t-t'$ erasures, for some $0 \le t' \le t$.
Since the $t$-error-correcting code over the finite field (e.g., BCH code) is capable of decoding $t'$ errors and $2(t-t') \ge t-t'$ erasures, the decoding process remains effective.
\end{remark}

\subsection{Reduced class codes}
In order to reduce the finite field size and hence lower the encoding/decoding complexity, we provide an alternative classification for $l \le 4$ with fewer classes, termed \emph{reduced classes}.

 We will describe necessary conditions for valid classifications, establish valid reduced classes in Theorem \ref{thm:valid class}, and finally present the LMPE correction code using reduced classes in Construction \ref{cnstr:reduced classes}.
 
Similar to Construction \ref{cnstr:remainder classes}, we first divide the probability vectors by $(2l+1)$ and obtain $(2l+1)^3$ possible remainder vectors. Then, we further partition them into $(2l+1)^2$ classes, named ``reduced classes", each with $(2l+1)$ remainder vectors. 
Such reduction of classes is possible because the classification in Framework \ref{cnstr:framework} only requires that the symbols in the same class cannot be converted to each other by an $l$-limited-magnitude probability error. 
Therefore, the valid classification is required to satisfy two conditions:

{\bf C1.} The difference between any two probability vectors in the same class is not an $l$-limited-magnitude probability error.

{\bf C2.} Each of the $(2l+1)^3$ remainder vectors is included in exactly one class.

Some notations and definitions are introduced below to clarify our presentation of the classification.
For a remainder vector $\mathbf{b}=(b_1,b_2,b_3,b_4)$ and an integer $i$, we denote
\begin{align*}
    i\mathbf{b}=(ib_1,ib_2,ib_3,ib_4) \mod (2l+1),
\end{align*}
as the scaled remainder vector. For two remainder vectors $\mathbf{b},\mathbf{c}$, by abuse of notation, we write
\begin{align*}
    \mathbf{b}+\mathbf{c} = \mathbf{b}+\mathbf{c} \mod (2l+1),
\end{align*}
as the sum remainder vector.
For magnitude $l$, a remainder vector $\mathbf{d}$ 
is said to be a \emph{remainder error pattern} if
\begin{align}
    \sum_{j\in[4]: d_j \in [0, l]} d_j = \sum_{j \in [4]:  d_j \in [l+1, 2l]} (2l+1-d_j)  \le l.\label{eq:error pattern}
\end{align}
If $\mathbf{d}$ is the difference between two probability vectors modulo $2l+1$, the first sum corresponds to the upward limited-magnitude probability errors, and the second sum corresponds to the downward errors. Note that above equation requires that the transmitted and the received remainder vectors have a Lee distance of $2l$, and the sum of the upward errors equals that of the downward errors.
Let $\mathbf{b}$ be a remainder vector where $b_1=1$, $\sum_{j=1}^{4}b_j \mod (2l+1) =0$, and $i\mathbf{b}$ is not a remainder error pattern for any $i \in [2l]$. Then $\mathbf{b}$ is called a \emph{critical vector}.

We provide an example of the classification of remainder vectors for limited-magnitude $l$ and $k \mod (2l+1) = 0$ in Table \ref{table:reduced classes}. In the table, there are $(2l+1)^2$ rows and $(2l+1)$ columns. Each row corresponds to a class. And each column shares the same first element of remainder vectors. Column 0 lists all remainder vectors where the first entry is $0$ and the sum is $0$ modulo $2l+1$. Remainder vectors in Column $i$ correspond to the remainder vectors in Column $0$ added by $i\mathbf{b}$, for $i \in [2l]$. Here, $\mathbf{b}$ is a critical vector. In general, for any $k \mod (2l+1) \neq 0$, the last entry in each remainder vector is added by $k$ modulo $2l+1$. For instance, if $k=14, l=1$, $(0,0,1,2l)$ in Class 1 and Column 0 becomes $(0,0,1,2l+k)  \mod 3  = (0,0,1,1)$.

\begin{table}[b]
    \centering
    \caption{Classification of remainder vectors for limited-magnitude $l$ and $k \mod (2l+1) = 0$. }
    \label{table:reduced classes}

\begin{tabular}{|c|c|c|c |c|} 
 \hline

 Class & Column $0$ & Column $1$ & ... & Column $2l$\\
 \hline
 0& $(0,0,0,0)$&$\mathbf{b}$&...&$2l \mathbf{b}$  \\ 
 \hline
1 & $(0,0,1,2l)$& $(0,0,1,2l)+\mathbf{b}$ &...& $(0,0,1,2l)+2l \mathbf{b}$  \\ 
 \hline

 ...&...&...&...&...\\

 \hline
$(2l+1)^2$ & $(0,2l,2l,2)$ & $(0,2l,2l,2)+\mathbf{b}$ &...& $(0,2l,2l,2)+2l \mathbf{b}$ \\
 \hline
\end{tabular}
\label{classes}
\end{table}

Next, we are ready to provide theorem states that the above classification is valid if $b$ is a critical vector.

\begin{theorem}\label{thm:valid class}
 If $\mathbf{b}$ is a critical vector, then the classification with general $k$ in Table \ref{table:reduced classes} is valid.
\end{theorem}
\begin{IEEEproof}
We need to prove that Conditions C1 and C2 are satisfied.
    
{\bf C1.} The remainder vectors in Class $i$ are in the form of $\mathbf{c}+i\mathbf{b}$, $0 \le i \le 2l$, where $\mathbf{c}$ is the vector in Column 0 of Table \ref{table:reduced classes}. Therefore, the difference between two symbols in a class modulo $2l+1$ must be $i\mathbf{b}$. Let $i\mathbf{b}=(d_1,d_2,d_3,d_4)$.
Assume C1 does not hold. Then there exist two probability vectors in the same class, such that their difference $\mathbf{e}=(e_1,e_2,e_3,e_4)$ is an $l$-limited-magnitude probability error. 
It can be checked that since $|e_j|\le l$ for $j\in [4]$, 
\begin{align}
    |e_j|= \begin{cases}
    d_j, & \text{ if } d_j \in [0,l],\\
    2l+1-d_j, & \text{ if } d_j \in [l+1,2l].
    \end{cases}
\end{align}
Therefore,
\begin{align}
    & \sum_{j\in[4]: d_j \in [0, l]} d_j + \sum_{j \in [4]:  d_j \in [l+1, 2l]} (2l+1-d_j)\\
   =& ~~~~ \sum_{j=1}^4 |e_j| \le 2l,
\end{align}
where the last inequality follows from Lemma \ref{lem:limited error}. Now there is a contradiction to the definition of the critical vector, where $i\mathbf{b}=(d_1,d_2,d_3,d_4)$ must not satisfy \eqref{eq:error pattern}.

{\bf C2.} 
Any vector in the $0$-th column sums to $k$ modulo $(2l+1)$.
Adding $i\mathbf{b}$ does not change the sum of the remainder vector modulo $(2l+1)$ because $\sum_{j=1}^{4}b_j \mod (2l+1) =0$. Therefore, all vectors listed in the table sum to $k$ modulo $(2l+1)$, as desired.
Since the $0$-th column lists all remainder vectors where the first entry is $0$, and the first entry of $\mathbf{b}$ is $1$, adding $i\mathbf{b}$ gives all remainder vectors where the first entry is $i$ in the $i$-th column, for $1 \le i \le 2l$. Hence, each remainder vector is listed in exactly one class.
\end{IEEEproof}

A critical vector $\mathbf{b}$ can be found by checking \eqref{eq:error pattern} for every $i\mathbf{b}$.
We list one critical vector for each $l$, $l \le 4$, in Table \ref{table:critical vector}. As the error magnitude $l$ increases, the number of remainder error patterns becomes larger. When $l>4$, the number of critical vectors becomes $0$. Hence, the classification method is suitable when $l \le 4$. 

\begin{table}
    \centering
    \caption{Critical vectors for $l\le 4$.}
    \label{table:critical vector}
    \begin{tabular}{|c|c|}
    \hline
    $l$ & Critical vector  \\
    \hline
    1 & $(1,1,1,0)$\\
    \hline
    2 & $(1,1,2,1)$\\
    \hline
    3 & $(1,2,3,1)$\\
    \hline
    4 & $(1,4,6,7)$\\
    \hline
\end{tabular}
\end{table}

Now we are ready to describe the code construction using the above classification.

\begin{construction}[Reduced classes, $l\le 4$]\label{cnstr:reduced classes}

{\bf Symbol classification:} Divide each probability vector by $(2l+1)$ to get the remainder vector based on Definition 4, and obtain $(2l+1)^2$ reduced remainder classes as Table \ref{table:reduced classes}.

{\bf First layer:} Apply a BCH code over $GF(q_1)$ to the reduced remainder classes. The code length is $n= q_1^w -1$, $q_1=(2l+1)^2$, and the code distance is $2t+1$. 

{\bf Second layer:} Apply a BCH code over $GF(q_2)$ to the first element of the remainder vector, which is shown as the first element in each column of Table \ref{table:reduced classes}. The code length is $n= q_2^w -1$, $q_2=2l+1$, and the code distance is $t+1$.
\end{construction}

\begin{theorem}
    Construction \ref{cnstr:reduced classes} can correct any $(l,t)$ LMPE.
\end{theorem}
\begin{IEEEproof}
    Due to the coding in the first layer, the decoder can find the locations of the erroneous symbols and their correct classes.
Given the received symbols, the limited-magnitude probability errors can be corrected if we know the correct remainder vectors as in Construction \ref{cnstr:remainder classes}. Furthermore, a remainder vector is uniquely determined by its first entry knowing the class index, because the first entry of the remainder vector in the $i$-th column equals $i$ in Table \ref{table:reduced classes}. Finally, the first entries of remainder vectors can be recovered by the code of distance $t+1$ in the second layer since we know the error locations (treating the $t$ errors as erasures).
\end{IEEEproof}

\subsection{Codes based on the improved Hamming code}
The following construction is an improvement of Construction \ref{cnstr:remainder classes} with Hamming code for $(l,t=1)$ LMPE, inspired by the work on efficient non-binary Hamming codes for limited-magnitude probability errors on integers \cite{Improved_haming}. The main idea is that for different transmitted remainder class indices in $GF((2l+1)^3)$, their erroneous class indices do not contain all possible elements in $GF((2l+1)^3)$. 
Thus, for the first layer of Construction \ref{cnstr:remainder classes}, an improved Hamming code over $GF((2l+1)^3)$ only needs to protect against some errors instead of all $(2l+1)^3$ possible errors. The improved code benefits from a higher code rate for the same redundancy $r$ compared to the Hamming code. Details of the parity check matrix construction of the improved Hamming code are in Appendix \ref{C}.

\begin{construction}[Improved Hamming code, $l, t=1$]\label{cnstr:improved Hamming}

{\bf Symbol classification:} Divide each probability vector by $(2l+1)$ to get the remainder vector based on Definition \ref{def: quo&rem}, and accordingly obtain $(2l+1)^3$ classes. 

{\bf First layer:} Apply a improved Hamming code constructed as in Appendix \ref{C} to the reduced remainder classes.

{\bf Second layer:} Not needed.

\end{construction}

We provide an example of the systematic parity check matrix of the improved Hamming code with $r=2$ redundancies is as below:
\begin{align}
    \begin{bmatrix}
     0&1&...&1&7&...&7&1&0\\ 
     7&1&...&26&0&...&26&0&1
    \end{bmatrix},
 \end{align} 
where the integers denote the exponent in the power representation of $GF(27)$, and the primitive polynomial $x^3+2x+1$ is used to represent elements in $GF(27)$. 
Thus, we get a $(56,54)$ Hamming code, which has a better rate than the original $(28,26)$ Hamming code.

Unless otherwise stated, we assume that $(2l+1)^3$ is a prime power. If $(2l+1)^3$ is not a prime power, our Construction \ref{cnstr:improved Hamming} remains effective. The encoding and decoding is similar to Remark \ref{remark: 4}.

\begin{remark}
According to the method described in Appendix \ref{C}, there can be multiple ways to generate the improved Hamming code, and the resulting code length and code rate can be different. Theorem \ref{thm:improved_Hamming_rate} in Appendix \ref{C} states that the optimal length $n= 2\frac{q^r-1}{q-1}$ of the improved Hamming code is attainable, if $q=(2l+1)^3$ is a prime power. Accordingly, the LMPE code in Construction \ref{cnstr:improved Hamming} has redundancy $\log_2 \left(\frac{(2l+1)^3-1}{2}n+1\right)$ in bits. 
\end{remark}

\subsection{Comparison}
In the following, the proposed code constructions are compared. As illustrative examples, we first consider the codes for $(l=1,t=1)$ LMPE. Then we compare the performance when there are $t$ errors. For the measurement criteria, we compare the field size which is related to the computational complexity, as well as the redundancy in bits with fixed codeword length $n$. 

The results are in Table \ref{table:redundancy} whose derivation is in Appendix \ref{app:redundancy}. In the table, naive Hamming code refers to the single-error correction code applied to the entire probability-vector symbol, requiring a field of size of at least $\binom{k+3}{3}$. When there are $t$ errors, we list codes using BCH codes whose maximum possible redundant bits are shown. The length of the Hamming code is assumed to be $n=\frac{q^r-1}{q-1}$, and that of the BCH code is $n=q^w-1$, where $q$ is the field size, $r$ is the number of redundant symbols, and $w$ is any integer. For BCH code, the redundancy satisfies $r \le 2t w$. Moreover, the redundancies in the table are approximated assuming that $\frac{k}{l}$ is large.  
 \begin{table*}[t]
 \small
\centering
\caption{Comparisons for the proposed constructions. The redundancy bits are calculated for $\frac{k}{l}$ sufficiently large.}
\label{table:redundancy}
\begin{tabular}{|c| c| c|c|}

 \hline

Method &Error &Field size &Redundancy in bits\\
\hline
Naive Hamming code& $l=1,t=1$& $\binom{k+1}{3}$ & $\log_2\left(\left(\binom{k+3}{3}-1\right)n+1\right)$ \\ 
\hline
Hamming code with remainder classes& $l=1,t=1$ & $27$ & $\log_2(26n+1)$\\ 
 \hline
Improved Hamming with remainder classes& $l=1,t=1$ & $27$ & $\log_2(13n+1)$\\
\hline
Hamming code with  reduced classes & $l=1,t=1$ & $9$ & $\log_2(8n+1)+\log_2(3)$\\ 
\hline
BCH code  with remainder classes& $l,t$ & $(2l+1)^3$ &$2t \log_2(n+1)$ \\
\hline
BCH code with reduced classes &$l\le 4,t$ & $(2l+1)^2$ & $3t \log_2(n+1)$\\
 \hline
 \end{tabular}
\end{table*}

The improved Hamming code has the least redundant bits when there is a single error ($t=1$) with limited magnitude $l=1$. 
For arbitrary $t$ errors and magnitude $l \leq 4$, BCH code with remainder classes has less redundancy than BCH code with reduced classes. The reason is that the maximum possible number of redundant bits in BCH code remains the same irrespective of the field size. However, the reduced classes require extra coding in the second layer, leading to a worse code rate. However, since the encoding/decoding operations are over smaller fields, the reduced classes benefit from lower complexity.

\begin{remark}
In Example \ref{example:1}, the number of possible information messages represented in a parity quotient vector is only $10$, which was to accommodate the least number of possible quotient vectors for a given remainder vector. 
It can be seen from Theorem \ref{prop: redundancy for general t} of Appendix \ref{app:redundancy} (in particular, the approximation step in \eqref{eq:54}) that when the resolution $\frac{k}{l}$ is sufficiently large, we can ignore the loss for the number of possible information messages represented by the parity quotient vectors, and only consider the redundancy in terms of the parity remainder vectors. Similar conclusions can be made for all the proposed constructions.
\end{remark}

\begin{remark}
Calculations in Theorem \ref{prop: redundancy for general t2} of Appendix \ref{app:redundancy} show that the number of redundant symbols for BCH codes is the same in both layers in Construction \ref{cnstr:reduced classes}. Hence, its code structure bears similarity to that of Construction \ref{cnstr:remainder classes}. Namely, the codeword can be divided into information symbols and parity symbols, where the remainder vectors in the parity symbols are redundant. See Figure \ref{fig:reduced classes} (a) of Appendix \ref{app:redundancy}. 
\end{remark}

\subsection{Asymptotic optimality}
While the constructions in the previous sections are applicable to finite $k$ and $n$, and can be selected according to the system's requirements, this section demonstrates that Construction \ref{cnstr:remainder classes} based on remainder classes yields the largest possible codes for LMPE in the asymptotic regime.

As in Appendix \ref{app:redundancy}, let $\mathcal{P}$ be the set of probability vectors with resolution $k$ as in \eqref{eq:def_P}, and $\mathcal{R}$ be the set of remainder vectors as in \eqref{eq:def_R} whose values are between $0$ and $2l$.

\begin{definition}\label{def:LMPE_rem}
The $(l,t)$ LMPE error word $\mathbf{e}=(\mathbf{e}_1,\mathbf{e}_2,\dots,\mathbf{e}_n)$ of the remainder vectors $\mathcal{R}$ is defined similar to Definition \ref{def:error}: 
\begin{align}
    &|\{i \in [n]: \mathbf{e}_i \neq \mathbf{0}\}| \le t,\\
    &\sum_{j \in [m]: e_{i,j}>0} e_{i,j} = -\sum_{j \in [m]: e_{i,j}<0} e_{i,j} \le l, \forall  i \in [n]. \label{eq:LMPE_rem}
\end{align}
Furthermore, given a remainder word $\mathbf{r}$, an LMPE word $\mathbf{e}$ must satisfy that every entry of $\mathbf{r+e}$ is between $0$ and $2l$. 
\end{definition}

It should be noted that the errors in Definition \ref{def:LMPE_rem} are different from the errors observed in Construction \ref{cnstr:remainder classes}. For example, when $l=1$, the probability vector $(3,3,3,3)$ can have a limited-magnitude probability error and becomes $(4,2,3,3)$. Correspondingly, the remainder vector changes from $\mathbf{r}=(0,0,0,0)$ to $\mathbf{r'}=(1,2,0,0)$. In Construction \ref{cnstr:remainder classes}, the error correction code on the remainder vectors needs to correct the remainder error $\mathbf{r}'-\mathbf{r}=(1,2,0,0)$. However, the condition in \eqref{eq:LMPE_rem} requires each value of the error to be between $-l$ and $l$, and hence $(1,2,0,0)$ is not considered an error for remainder vector $(0,0,0,0)$. In fact, for the remainder vector $(0,0,0,0)$, there is no error according to Definition \ref{def:LMPE_rem}. The condition in \eqref{eq:LMPE_rem} will be demonstrated to be appropriate in Theorem \ref{thm:optimal}.

\begin{definition}\label{def:sizes}
Let the resolution be $k$.
Define the sizes of three kinds of codes as below: 
\begin{itemize}
    \item $A_{prob, LMPE}(n,k,t,l) $: the size of the largest length-$n$ code over the probability vectors $\mathcal{P}$ that corrects $(l,t)$ LMPE.
    \item $A_{rem, LMPE}(n,k,t,l) $: the size of largest length-$n$ code over the remainder vectors $\mathcal{R}$ that corrects $(l,t)$ LMPE. 
    \item $A_{rem,arbitrary}(n,k,t,l) $: the size of the largest length-$n$ code over the remainder vectors $\mathcal{R}$ that corrects $t$ arbitrary errors.
\end{itemize}
\end{definition}

{\bf Informal description of the optimality result.} In this section, Construction \ref{cnstr:remainder classes} refers to the construction whose first layer uses an optimal code over $\mathcal{R}$ for arbitrary errors with size $A_{rem,arbitrary}(n,k,t,l)$. Note that the code in the first layer is not tailored for limited-magnitude probability errors, potentially resulting in a suboptimal overall code size. However, our optimality result states that an optimal code in the first layer is sufficient to ensure an asymptotic optimal LMPE code over the probability vectors.

{\bf Proof outline.} The main idea of our optimality proof is as follows.
A lower bound on the code size of Construction \ref{cnstr:remainder classes}, and hence a lower bound on $A_{prob, LMPE}(n,k,t,l)$, is shown in Theorem \ref{thm:lower_bound}, which is based on $A_{rem,arbitrary}(n,k,t,l)$. 
The upper bound on $A_{prob, LMPE}(n,k,t,l)$ is obtained in Theorem \ref{thm:optimal} based on $A_{rem, LMPE}(n,k,t,l)$, which is further upper bounded by a function of $A_{rem,arbitrary}(n,k,t,l)$ due to Lemma \ref{lm: lmnlm}.
The key method for the upper bound is to partition the optimal LMPE correction code over $\mathcal{P}$, and obtain an LMPE correction code over $\mathcal{R}$.
Observing that the lower and the upper bounds are asymptotically identical, we conclude that Construction \ref{cnstr:remainder classes} is asymptotically optimal.

The optimality proof is inspired by \cite{Cassuto_flash}, but several steps are different for probability vectors. For example, when we partition the optimal LMPE correction code over $\mathcal{P}$ in Theorem \ref{thm:optimal}, the sizes of the partitions need to be carefully bounded because each quotient vector corresponds to different numbers of remainder vectors. For another example, in Lemma \ref{lm: lmnlm}, the sizes $A_{rem, LMPE}(n,k,t,l)$ and $A_{rem,arbitrary}(n,k,t,l)$ are related in a different way from \cite{Cassuto_flash}.

\begin{theorem}\label{thm:lower_bound}
Fix $n,t,k,l$.
Let $s=\left\lfloor\frac{k}{2l+1}\right\rfloor$. The size of the code $\mathcal{C}$ of Construction \ref{cnstr:remainder classes} is bounded by the following inequalities:
\begin{align}
 \binom{\max\{s,3\}}{3} ^n A_{rem,arbitrary}(n,k,t,l) \leq |\mathcal{C}| \leq \binom{s+3}{3} ^n  A_{rem,arbitrary}(n,k,t,l) . \label{lb}
\end{align} 
\end{theorem}
\begin{IEEEproof}
Let $\mathcal{C}'$ be the largest length-$n$ code over $\mathcal{R}$ that corrects $t$ arbitrary errors whose size is $A_{rem,arbitrary}(n,k,t,l) $.
By Construction \ref{cnstr:remainder classes}, a valid codeword of $\mathcal{C}$ is obtained by using a codeword of $\mathcal{C}'$ as the remainder vectors added by some quotient vectors times $2l+1$. The number of possible quotient vectors for a fixed remainder vector is denoted as $Q'$, as mentioned in Appendix \ref{app:redundancy}. From \eqref{eq:26}, $\binom{\max\{s,3\}}{3} \le Q' \le  \binom{s+3}{3}$, and the statement holds. 
\end{IEEEproof}

The gap between $A_{rem, LMPE}(n,k,t,l) $ and $A_{rem,arbitrary}(n,k,t,l) $ can be bounded by using the following lemma. The idea is that, for the alphabet $\mathcal{R}$, we take a subset from a code for limited-magnitude probability errors, and obtain a code for arbitrary errors.  
\begin{lemma}\label{lm: lmnlm}
Fix $n,t,k,l$.
Let $f(l) = |\mathcal{R}| = (2l+1)^3$. The following inequality is satisfied:
$$A_{rem,arbitrary}(n,k,t,l)  \geq \frac{1}{2t n^{2t} f(l)^{2t}} A_{rem, LMPE}(n,k,t,l). $$ 
\end{lemma}
\begin{IEEEproof}
Let $\mathcal{C}'$ be an $(l,t)$ LMPE code of length $n$ over $\mathcal{R}$ whose size is $A_{rem, LMPE}(n,k,t,l) $.  
Any two codewords in  $\mathcal{C}'$ have a Hamming distance of at least $1$. The number of words (and hence an upper bound on the codewords) that are at Hamming distance between $1$ and $2t$ from a codeword of a $\mathcal{C}'$ is at most
 
\begin{align*}
    \sum_{i=1}^{2t}\binom{n}{i}f(l)^i  < 2t n^{2t} f(l)^{2t}.
\end{align*}


We claim that there exists a code $\mathcal{C}''$  of length $n$ over $\mathcal{R}$ that corrects $t$ arbitrary errors, such that any two codewords have a Hamming distance of at least $2t+1$. 
In particular, keeping at least $1/\left(2t n^{2t} f(l)^{2t}\right)$ of the codewords from $\mathcal{C}'$ yields a code $\mathcal{C}''$ of Hamming distance at least $2t+1$. Thus,
\begin{align*}
         A_{rem,arbitrary}(n,k,t,l)   \ge |\mathcal{C}''| \ge  \frac{1}{2t n^{2t} f(l)^{2t}} A_{rem, LMPE}(n,k,t,l).
\end{align*}
The proof is completed.
\end{IEEEproof}

In the following theorem, 
we consider a sequence of codes with parameters $(n,k,l,t)$,
such that the block length $n \to \infty$, the resolution $k \to \infty$, 
the number of errors $t = o(n/\log n)$, and the error magnitude $l = O(n)$. In particular, $l$ grows at most linearly with $n$ (but it is allowed that $l$ grows slower than  $n$). 
We show the asymptotic optimality of Construction \ref{cnstr:remainder classes} for this sequence.

\begin{theorem}\label{thm:optimal}
Given a sequence of optimal $(l,t)$ LMPE code over $\mathcal{R}$ of size $A_{rem, LMPE}(n,k,t,l)$, where $n \to \infty, k \to \infty, t = o(n/ \log n), l = O(n)$,
Construction \ref{cnstr:remainder classes} generates a sequence of asymptotically optimal $(l,t)$ LMPE code over $\mathcal{P}$. 
\end{theorem}
\begin{IEEEproof}
Partition the probability vectors $\mathcal{P}$ into ``quotient sets'' such that each set shares the same quotient vector after dividing by $(2l+1)$. Thus, the number of quotient sets equals to the total number of quotients, denoted as $Q$ in Appendix \ref{app:redundancy}. By  \eqref{eq:27}\eqref{eq:28}, it satisfies
\begin{align}
Q \le \sum_{j=s-3}^{s} \binom{j+3}{3} \le 4 \binom{s+3}{3} \le \frac{2(s+3)^3}{3}.   \label{eq:35}
\end{align}

For $\mathbf{x}\in \mathcal{P}^n$, let $\mathcal{B}_{l,t}(\mathbf{x}, \mathcal{P})$ be the error ball of all words in $\mathcal{P}^n$ that differ from $\mathbf{x}$ by an $(l,t)$ LMPE. Similarly, for $\mathbf{r} \in \mathcal{R}^n$, define the error ball $\mathcal{B}_{l,t}(\mathbf{r}, \mathcal{R})$ to be the words in $\mathcal{R}^n$ that differ from   $\mathbf{r}$ by an $(l,t)$ LMPE. By Definitions \ref{def:error} and \ref{def:LMPE_rem}, the LMPE of $\mathcal{P}$ and the LMPE of $\mathcal{R}$ are defined in the same way, except that an LMPE of $\mathcal{R}$ added to the remainder vector must still have entries between $0$ and $2l$. 

Consider an optimal code $\mathcal{C}$ over $\mathcal{P}$ correcting $(l,t)$ LMPE, whose size is $A_{prob, LMPE}(n,k,t,l)$. Then there exists one subset of $\mathcal{C}$, whose quotient vectors are all the same
and the size is at least $A_{prob, LMPE}(n,k,t,l)/Q^n$. Since the code corrects $t$ errors, for any $\mathbf{x},\mathbf{y}$ in this subset,

\begin{align*}
    \mathcal{B}_{l,t}(\mathbf{x},\mathcal{P}) \cap \mathcal{B}_{l,t}(\mathbf{y},\mathcal{P}) = \emptyset.
\end{align*}
Since both $\mathbf{x}$ and $\mathbf{y}$ have the same quotients of length $n$, say $\mathbf{q}$, we get
\begin{align*}
    \mathcal{B}_{l,t}(\mathbf{x}-\mathbf{q},\mathcal{R}) \cap \mathcal{B}_{l,t}(\mathbf{y}-\mathbf{q},\mathcal{R}) = \emptyset.
\end{align*}
Therefore, we get a code with size at least $A_{prob, LMPE}(n,k,t,l)/Q^n$ that corrects $(l,t)$ LMPE over $\mathcal{R}$. By Definition \ref{def:sizes},
\begin{align}
    A_{rem, LMPE}(n,k,t,l)  \geq \frac{1}{Q^n} A_{prob, LMPE}(n,k,t,l) . \label{ub}
\end{align}

Combining the lower bound \eqref{lb} and upper bound \eqref{ub}, we obtain
\begin{align}
   &\binom{\max( s,3)}{3}^n A_{rem,arbitrary}(n,k,t,l)   \\
   &\leq A_{prob, LMPE}(n,k,t,l)  \\
   &\leq  Q^n A_{rem, LMPE}(n,k,t,l) , \label{lbub}
\end{align}
where $s = \left\lfloor \frac{k}{2l+1} \right\rfloor$.

Combining Lemma \ref{lm: lmnlm} and \eqref{lbub}, 
\begin{align*}
    &\binom{\max\{s,3\}}{3} ^n A_{rem,arbitrary}(n,k,t,l) \\
    \leq & A_{prob, LMPE}(n,k,t,l)  \\
    \leq & 
    \left(2t n^{2t} f(l)^{2t} \right) Q^n A_{rem,arbitrary}(n,k,t,l) .
\end{align*}
Taking the logarithm base $q=\binom{k+3}{3}$, dividing by $n$, and taking the limit $n \to \infty$, we obtain bounds on the code rate:
\begin{align}
 &\lim_{n \to \infty}  {\log_q \binom{\max\{s,3\}}{3} } + \lim_{n \to \infty} \frac{\log_q A_{rem,arbitrary}(n,k,t,l) }{n}\label{eq:38}\\
   \le & \lim_{n \to \infty} \frac{\log_q A_{prob, LMPE}(n,k,t,l) }{n} \\
   \le & \lim_{n \to \infty} \log_q Q + \lim_{n \to \infty} \frac{\log_q A_{rem,arbitrary}(n,k,t,l) }{n} , \label{eq:40}
\end{align}
where we used $t = o(n/\log n)$, $l = O(n)$. 
The gap between the lower bound \eqref{eq:38} and the upper bound \eqref{eq:40} is:
\begin{align}
& \log_q Q - \log_q  \binom{\max\{s,3\}}{3} \label{eq:44}.
\end{align}

Assume $k \to \infty$, and thus $q = \binom{k+3}{3} \to \infty$. 
We consider two cases. 

Case I. When $s=\left\lfloor \frac{k}{2l+1} \right\rfloor=0, 1$ or $2$, the number of possible quotient sets is $Q=1$, 
$Q=\binom{3}{3}+\binom{4}{3}=5$ or $Q=\binom{3}{3}+\binom{4}{3}+\binom{5}{3}=15$, respectively. The gap in \eqref{eq:44} simplifies to $\log_q Q \le \log_q 15$. 

Case II. When $s \ge 3$, by \eqref{eq:35}, the gap in \eqref{eq:44} becomes
\begin{align*}
& \log_q \frac{Q}{\binom{s}{3}} \\
\le & \log_q \frac{4 (s+3)^3/6 }{(s-2)^3/6} \\
= & \log_q 4 + 3 \log_q \left(1+ \frac{5}{s-2}\right)\\
\le & \log_q 4 + 3 \log_q \left(1+ \frac{5}{3-2}\right)\\
=& \log_q 864.
\end{align*}

Combining both cases, the gap is at most $\log_q 864$, which tends to $0$ as $k \to \infty$.
Therefore, the sequence of codes generated by Construction \ref{cnstr:remainder classes} is asymptotically optimal.
%

\end{IEEEproof}

\subsection{Systematic LMPE codes }
 
In order to facilitate easy access to the information contained within the probability codewords, we introduce systematic LMPE codes in this subsection. The systematic code is based upon Construction \ref{cnstr:remainder classes}. Recall that the remainder vector is viewed as a finite field element of size at least $(2l+1)^3$. To construct systematic codes, Gray codes are introduced to represent the parity symbols (probability vectors), such that a $2l$-limited-magnitude probability error only results in 1 finite field error. Then we show that an $(l,t)$ LMPE correction code over the finite field is sufficient to protect the messages.

The structure of the systematic code is inspired by \cite{Cassuto_flash}. Different from \cite{Cassuto_flash}, the mapping between remainder vectors and finite field elements needs to be defined similarly to Construction \ref{cnstr:remainder classes}. Additionally, Gray mapping between the probability vectors and the finite field elements must be established to achieve the desired error correction capability.

We first define Gray code, which is an important component of our construction. Then, we present the systemic code in Construction \ref{cnstr:systematic} and its correctness in Theorem \ref{thm:systematic}, followed by an algorithm to find a Gray code in Algorithm \ref{alg:gray mapping}.

\begin{definition}[Gray code]\label{def:gray}
Let $g$ be an integer.
Every vector over $GF(q)$ of length $g$ is called a \emph{Gray codeword}.
A \emph{Gray code} maps each Gray codeword to a probability vector of resolution $k$ and satisfies the condition below. 
If two probability vectors differ by a $2l$-limited-magnitude probability error and their Gray codewords both exist, the Hamming distance of their Gray codewords must be $1$. 
\end{definition}

{\bf Notation.} To differentiate probability vectors and Gray codewords, we say a probability vector is a \emph{symbol} and an element in $GF(q)$ is a \emph{digit}. We add subscripts to the notations, such as $\mathbf{a}_{prob}$ for a probability vector, and $\mathbf{a}_{gray}$ for a Gray codeword. 

In Definition \ref{def:gray}, every $g$-digits over $GF(q)$ is mapped to a probability vector, but not every probability vector is mapped to a $g$-digit codeword over $GF(q)$.

\begin{figure}
    \centering
    \includegraphics[width=0.8\linewidth]{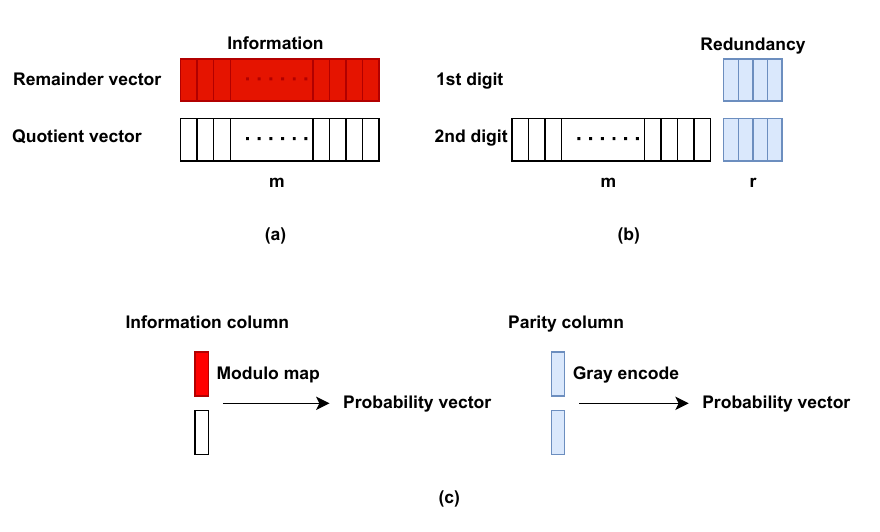}
    \caption{Encoding procedure for a systematic code. The Gray codeword consists of $g=2$ digits. (a) The information symbols are divided into two parts: remainder vector and quotient vector by the modulo operation. (b) The parities over the finite field are generated based on remainder vector, and every $g=2$ finite field digits are placed in one column. (c) The information symbols are formed by the modulo operation, and the parity symbols are formed based on the Gray code. }
    \label{fig:systematic}

\end{figure}

We construct a systematic $(l,t)$ LMPE code with $m$ information symbols and $r$ parity symbols.
Let $\mathcal{C}$ be a systematic $t$-error correction code over a finite field $GF(q)$ of size $q \leq (2l+1)^3$.  Assume there is a Gray code of length $g$ over $GF(q)$. Suppose $\mathcal{C}$ has $m$ information symbols, and $g\cdot r$ parity symbols.
If the number of parities is not a multiple of $g$, we pad $0$'s to form $gr$ parities. 
Each information symbol of the LMPE code can be divided into the quotient vector and the remainder vector through dividing by $(2l+1)$, shown in the same column in Fig. \ref{fig:systematic} (a). 
The remainder vector has $(2l+1)^3$ possibilities and is mapped to $GF(q)$. Using the remainder vectors as the information part of $\mathcal{C}$, we obtain $gr$ parities over $GF(q)$. Next, the $gr$ parities are placed in $r$ columns, each column containing $g$ digits over $GF(q)$ (Fig. \ref{fig:systematic} (b)). Finally, the $m+r$ columns are mapped to probability vectors, shown in Fig. \ref{fig:systematic} (c). The information column is mapped back to a probability vector from the remainder vector and quotient vector. The parity column (with $g$ digits over $GF(q)$) is mapped to the probability vector using a Gray code. 

\begin{construction}[Systematic $(l,t)$ LMPE code]\label{cnstr:systematic}

{\bf Symbol classification:} Divide each probability vector by $(2l+1)$ to get the remainder vector based on Definition \ref{def: quo&rem}, and accordingly obtain $(2l+1)^3$ classes. 

{\bf First layer:} Apply a BCH code over $GF(q)$ to the remainders, whose length is $n= q^w -1$, $q$ is prime power, $q \leq(2l+1)^3$ and distance is $2t+1$ and apply a Gray code in $GF(q)$ to the parities remainder vectors as shown as Fig. \ref{fig:systematic}. 

{\bf Second layer:} not needed.

\end{construction}

\begin{theorem}\label{thm:systematic}
    Construction \ref{cnstr:systematic} is a systematic $(l,t)$ LMPE correction code with rate $\frac{m}{m+r}$.
\end{theorem}
\begin{IEEEproof}
    It is obvious that the construction is systematic and the rate is $\frac{m}{m+r}$.
    We only need to prove that it can correct any $(l,t)$ LMPE.
    
    To decode, we first obtain the corrupted codeword of $\mathcal{C}$ over $GF(q)$ as follows. 
    (i) The remainder vector of each information column is mapped to one digit over $GF(q)$. An $l$-limited-magnitude probability error in the information column corresponds to an error over $GF(q)$ in the remainder vector as explained in Construction \ref{cnstr:remainder classes}. 
    (ii) For a parity column, 
    let the received parity probability vector be $\mathbf{b}_{prob}$. Note that $\mathbf{b}_{prob}$ may not correspond to a Gray codeword by Definition \ref{def:gray}. 
    Choose the Gray codeword $\hat{\mathbf{a}}_{gray}$ corresponding to any probability vector $\hat{\mathbf{a}}_{prob}$ within the $l$-limited-magnitude probability error ball  of $\mathbf{b}_{prob}$. 
    Such a probability vector $\hat{\mathbf{a}}_{prob}$ exists (it can be the original stored parity probability vector). Moreover, $\hat{\mathbf{a}}_{prob}$ differs from the original stored probability vector by at most $2l$ LMPE,  
    because the original vector is a $l$ LMPE from the received vector, and the received vector is another $l$ LMPE from the decoded vector. 
    This also leads to at most one digit error over $GF(q)$ due to Gray mapping.
    
Since there are at most $t$ LMPEs in the probability vectors, there will be at most $t$ errors over $GF(q)$.
Thus the decoder of $\mathcal{C}$ will successfully recover the $m+gr$ digits over $GF(q)$. The information columns are recovered similarly to Construction \ref{cnstr:remainder classes} and the parity columns are recovered by Gray mapping.
\end{IEEEproof}

\begin{remark}
Unless otherwise stated, we assume that $(2l+1)^3$ is a prime power. If $(2l+1)^3$ is not a prime power, our Construction \ref{cnstr:systematic} remains effective.

For encoding/decoding, it is similar to Remark \ref{remark: 4}, with an additional requirement for Gray mapping.
When the field size $q$ is smaller than $(2l+1)^3$, we map the finite field elements to a subset of the remainder vectors.
We must add constraints to Algorithm \ref{alg:gray mapping} to ensure that all probability vectors in the Gray mapping have remainder vectors within the selected subset. This can be achieved by increasing the value of $k$.

\end{remark}

\begin{example}[$(l=1,t=3)$ systematic LMPE]
For $g=2$ digits, we can choose resolution $k=19$ (see Algorithm \ref{alg:gray mapping}) and alphabet size $\binom{k+3}{3}=1540$. Let $\mathcal{C}$ be the BCH code over $GF(27)$ with $m=16,gr=16$. 
The systematic code rate is $m/(m+r)=0.667$. On the contrary, Construction \ref{cnstr:remainder classes} gives a non-systematic code of rate $0.750$.
\end{example}

The searching algorithm for Gray mapping is shown in Algorithm \ref{alg:gray mapping}.
The algorithm maintains probability-vector mappings in a list, denoted by $M=[M_2,M_2,\dots]$. 
Note that by Definition \ref{def:gray}, every Gray codeword must be associated with a probability vector, so $M$ should be of size $q^g$ by the end of the algorithm.
Denote $M_i=\{\mathbf{a}_{prob,i} \to \mathbf{a}_{gray,i}\}$, for any $i=1,2,\dots$. We traverse the probability vectors by breadth-first search (Lines \ref{line4} -- \ref{line7}). Starting from an initial probability vector $\mathbf{a}_{prob,1}$, we visit all probability vectors in its \emph{error ball} with $2l$-limited-magnitude probability errors, while adding \emph{valid} probability-vector mappings to the list $M$. Then, we continue to visit the error ball of the second probability vector $\mathbf{a}_{prob,2}$, and so on.
Here, the new mapping $\{\mathbf{b}_{prob} \to \mathbf{b}_{gray} \}$ is defined to be \emph{valid}  if for any $\{\mathbf{c}_{prob} \to \mathbf{c}_{gray} \}$ in the list $M$ such that $\mathbf{c}_{prob}$ and $\mathbf{b}_{prob}$ differ by a $2l$-limited-magnitude probability error, $\mathbf{c}_{gray}$ and $\mathbf{b}_{gray}$ differ by $1$ digit.
In addition, the list $V$ keeps track of all visited probability vectors to avoid repeated computation.
If all probability vectors have been visited, but $M$ does not contain $q^g$ pairs, the Gray mapping search fails (Line \ref{line10}). Once the size of $M$ reaches $q^g$, the algorithm succeeds (Line \ref{line12}.)

\begin{algorithm}
\caption{Gray mapping between probability vectors and $g$-digit codewords over $GF(q)$. The parameters satisfy $q \leq (2l+1)^3$, $q^{g} \le \binom{k+3}{3}$.}
\label{alg:gray mapping}
\KwIn{The resolution $k$, the error magnitude $2l$, the field size $q$, the codeword length $g$.}
\KwOut{Gray mapping list $M=[M_1,M_2,\dots,M_{q^g}]$.} 
\tcp{Denote $M_i= \{\mathbf{a}_{prob,i} \to \mathbf{a}_{gray,i}\}$ for each $i$.}
Initialize $M$: $M = \emptyset$\; \label{line1}
\tcp{$V$ is the list of visited probability vectors.}
Initialize $V$: $V = \emptyset$\;
Initialize $i=1$\;
\While {$|M| < q^g$ }{
    \If{$M_i = \emptyset$\label{line:empty1}}{
        Set $ \mathbf{a}_{prob,i}$ to be any probability vector not in $V$\ and $\mathbf{a}_{gray,i}$ to be any Gray codeword not in $M$\;
        Add $M_i$ = \{$ \mathbf{a}_{prob,i} \to \mathbf{a}_{gray,i} \}$  to $M$\;
        Add $\mathbf{a}_{prob,i}$ to $V$\; \label{line:empty2}
    }


 \For{$\mathbf{b}_{prob} \in$ \text{the error ball with magnitude $2l$ of} $\mathbf{a}_{prob,i}$\label{line4}}{ 
\If{ $\mathbf{b}_{prob}$  $\notin$ $V$ and $\{\mathbf{b}_{prob} \to \mathbf{b}_{gray}\}$ is valid \label{line5}}
{
         append $\{ \mathbf{b}_{prob} \to \mathbf{b}_{gray}\} $ to $M$\; 
         
            }
Append $ \mathbf{b}_{prob}$ to $V$\; \label{line7}

            }

\If{$|V|= \binom{k+3}{3}$ and $|M| < q^g$}{
return ``$Fail$'' \label{line10}\;}
$i= i+1$\;
            }
return $M$. \label{line12}
\end{algorithm}

We show next that, for any fixed $l, g$ and $q\ge (2l+1)^3$, when the resolution $k$ is large enough,  Algorithm \ref{alg:gray mapping} can find a Gray mapping successfully, and hence a systematic LMPE code exits. To that end, we introduce the following notation. Substituting $l'=2l$ in \eqref{eq:E1}, we obtain an upper bound on the number of possible $2l$-limited-magnitude errors in a probability vector, denoted by $E_{2l}$: 
\begin{align*}
    E_{2l}=\frac{80}{3}l^3+20l^2+\frac{22}{3}l + 1.
\end{align*}

\begin{theorem}\label{thm:gray_exist}
    Fix $l, g$ and $q\le (2l+1)^3$. 
    When the resolution $k$ satisfies $\binom{k+3}{3} \ge q^g E_{2l}$, Algorithm \ref{alg:gray mapping} is guaranteed to find a Gray mapping. 
\end{theorem}
\begin{IEEEproof}
    In the worst-case scenario of running the breadth-first search (Lines \ref{line4} -- \ref{line7} of Algorithm \ref{alg:gray mapping}), no probability vectors in the $2l$-limited-magnitude error ball are added to the mapping list. Then Lines \ref{line:empty1} -- \ref{line:empty2} will add a probability vector outside the visited error balls to the mapping list. The worst-case process continues by adding one probability vector to the mapping list and excluding all other probability vectors in its error ball whose size is at most $E_{2l}$. If the total number of probability vectors $\binom{k+3}{3}$ is at lest $q^g E_{2l}$,  the algorithm is guaranteed to find the Gray mapping list of size $q^g$.
\end{IEEEproof}

\begin{figure*}[ht]
\centering
\begin{subfigure}{.5\textwidth}
  \centering
  \includegraphics[width=0.9\linewidth]{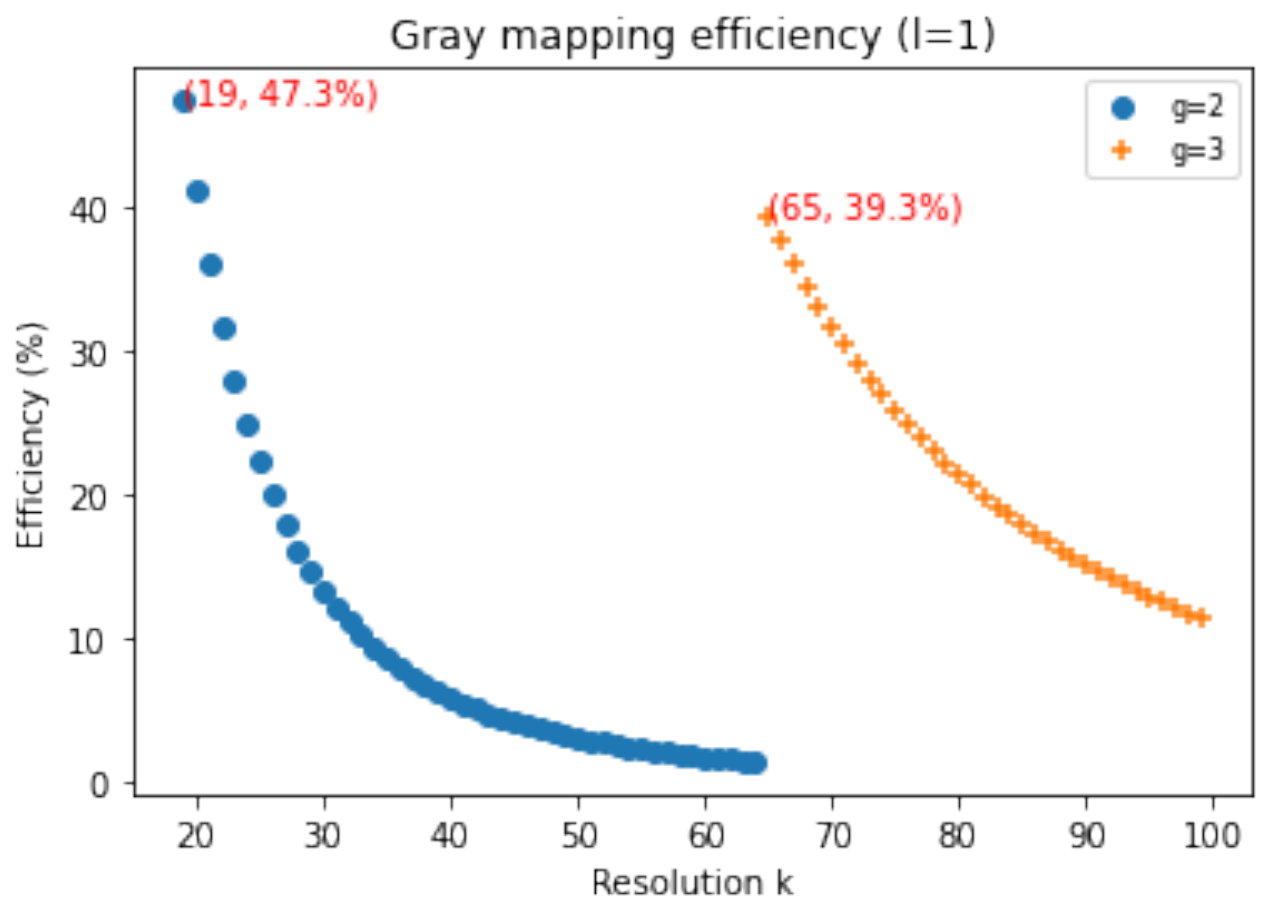}
  \caption{$l=1$ }
  \label{fig:sub1}
\end{subfigure}%
\begin{subfigure}{.5\textwidth}
  \centering
  \includegraphics[width=0.9\linewidth]{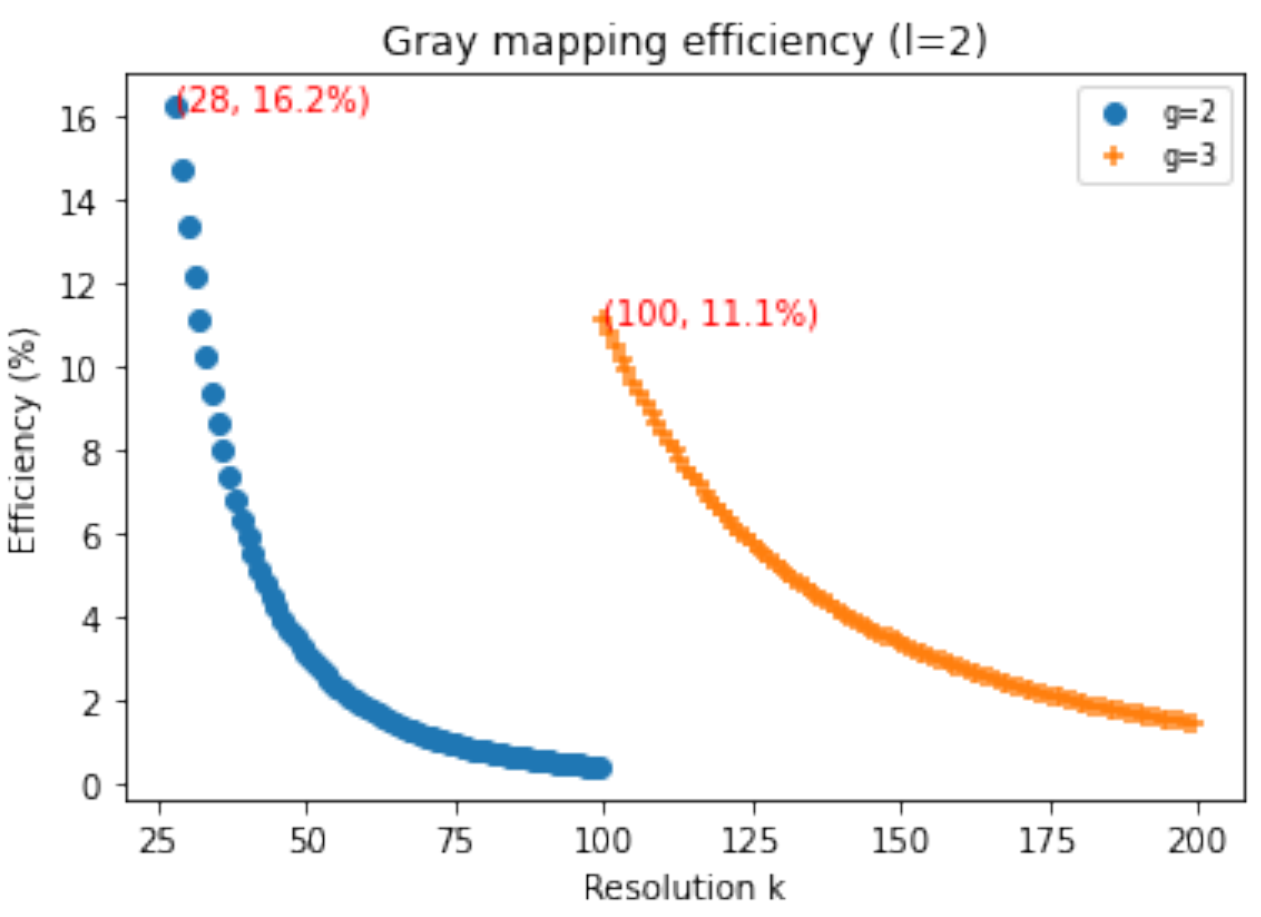}
  \caption{$l=2$}
  \label{fig:sub2}
\end{subfigure}
\caption{Gray mapping efficiency. For fixed $g,l$ and $q=(2l+1)^3$, the smallest $k$ found by Algorithm \ref{alg:gray mapping} is shown as the first number in the parenthesis. The second number in the parenthesis means the efficiency.}
\label{fig:Gray mapping}
\end{figure*}

The following theorem states that if a Gray code exists for some resolution, then it exists for all larger resolutions.

\begin{theorem}\label{thm:gray}
Let the number of Gray codeword digits $g$, the field size $q$,  and the error magnitude $l$ be fixed.
If there is a Gray mapping for resolution $k_1$, then there exists a Gray mapping for resolution $k_2$, for any $k_2>k_1$.  
\end{theorem}
\begin{IEEEproof}
Assume the Gray mapping exists for $k_1$. The mapping  for resolution $k_2$ is constructed as follows: for any $\{\mathbf{a}_{prob} \to \mathbf{a}_{gray}\}$ with resolution $k_1$, assign $\{\mathbf{a}'_{prob} \to \mathbf{a}_{gray}\}$ with resolution $k_2$,
where $\mathbf{a}'_{prob}$ is obtained by adding $k_2-k_1$ to the last element of $\mathbf{a}_{prob}$.
Next, we show that the constructed mapping for resolution $k_2$ is a Gray mapping.
For any $\{\mathbf{a}'_{prob} \to \mathbf{a}_{gray}\}$ and $\{\mathbf{b}'_{prob} \to \mathbf{b}_{gray}\}$ with resolution $k_2$, if $\mathbf{a}'_{prob}$ and $\mathbf{b}'_{prob}$ differ by a $2l$-limited-magnitude probability error, subtracting $k_2-k_1$ from the last element,  $\mathbf{a}_{prob}$ and $\mathbf{b}_{prob}$ also differ by a $2l$-limited-magnitude probability error. Hence, $\mathbf{a}_{gray}$ and $\mathbf{b}_{gray}$ differ by $1$ digit. 
\end{IEEEproof}

By Definition \ref{def:gray}, the resolution $k$, the finite field size $q$, and the number of digits $g$ must satisfy
\begin{align}
    q^{g} \leq \binom{k+3}{3} . \label{eq:gray_k}
\end{align}
For given $l,g,q$, let $k_0$ be the smallest resolution satisfying \eqref{eq:gray_k}.
By Theorem \ref{thm:gray}, we can run Algorithm \ref{alg:gray mapping} for $k=k_0$, and increase the resolution until it finds a Gray mapping successfully for some $k=k_1$. The existence of $k_1$ is ensured by Theorem \ref{thm:gray_exist}.
The constructive proof in Theorem \ref{thm:gray} provides a Gray code for any $k \ge k_1$. However, our method does not exclude possible Gray mapping for a smaller resolution. For example, the ordering of visiting the probability vectors, and the assignment of their Gray codewords can affect the success of Algorithm \ref{alg:gray mapping}. 
Next we analyze the \emph{efficiency} of the Gray mapping, defined as $q^g/\binom{k+3}{3}$. 
Fig. \ref{fig:Gray mapping} shows the smallest resolution $k$ found by Algorithm \ref{alg:gray mapping} for fixed $l,g,q$, and the change of efficiency as the resolution grows. For example, when $l=1$, $g=2$, the smallest $k$ is $19$; and the efficiency decreases from $47.3\%$ to $1.5\%$ when $k$ grows from $19$ to $64$. 

\begin{table*}[!]
\centering
\caption{Code rates for systematic and non-systematic codes. In each entry, the first value is the rate of the systematic code by Construction \ref{cnstr:systematic} and the second value is the rate of the non-systematic code by Construction \ref{cnstr:remainder classes}.}
\label{table:comparsion sys}
\begin{tabular}{|c|c|c|c|c|}
\hline
\diagbox{$k, l, g$}{$(n, m)$} & $(n=31,m=16) $& $(n=31, m=21)$ & $(n=63,m=51)$ & $(n=63, m=57)$\\
 \hline
$k=19$, $l=1, g=2$& 0.667 / 0.750  & 0.808 / 0.834 & 0.895 / 0.902 &  0.950 / 0.951 \\
\hline
$k=65$, $l=1, g=3$& 0.762 / 0.844 & 0.840 / 0.896  & 0.927 / 0.939 & 0.966 / 0.969\\ 
 \hline
$k=28$, $l=2, g=2$& 0.667 / 0.688 & 0.808 / 0.792 & 0.895 / 0.877 & 0.950 / 0.939 \\
\hline
$k=100$, $l=2, g=3$&0.762 / 0.798  & 0.840 / 0.902 & 0.927 / 0.942 & 0.966 / 0.960\\ 
 \hline
 \end{tabular}
\end{table*}

Examples based on BCH codes are listed in Table \ref{table:comparsion sys}.  
Some interesting observations can be made:
\begin{enumerate}
    \item Let $n,m$ be fixed. The systematic code rate $\frac{m}{m+r}$, where $r=\lceil \frac{n-m}{g} \rceil$, increases with $g$. For instance, $g=3$ in Rows 2 and 4 results in higher systematic code rates than $g=2$ in Rows 1 and 3.    
    However, the rate does not depend on $k$ or $l$ when $g$ is fixed. 
    For example, the systematic code rates in Row 1 ($k=19$, $l=1, g=2$) and Row 3 ($k=28$, $l=2, g=2$) are the same, even though the latter case has a lower Gray mapping efficiency as shown in Fig. \ref{fig:Gray mapping}. Similarly, Row 2 and Row 4 have the same systematic rates. On the other hand, for non-systematic codes, $k$ and $l$ both affect the rate.

    \item The rate of the systematic code is larger than the non-systematic code for fixed $n,m,k,l,g$ in some situations.
    See, for example, Row 3 ($k=28$, $l=2, g=2$), column 3 ($n=31,m=21$), column 4 ($n=63,m=51$) or column 5 ($n=63,m=57$). Notice that the non-systematic code rate depends on $k,l$, as explained in Theorem \ref{thm: non_sys} of Appendix \ref{app:redundancy}. In particular, the non-systematic code rate can be lower than that of the systematic code when $s_{\min}$ in Equation \eqref{eq:110} is small.
    
\end{enumerate}

For $k /(2l+1) \gg 1$, we compare the rates of the non-systematic code and the systematic code.
For simplicity, assume $g$ is a divisor of $n-m$ and $q = (2l+1)^3$. The code rate for the systematic code can be represented as 
$R_{s}=1-\frac{r}{m+r}$.
For the non-systematic code, the number of parities is $n-m=gr$, and the code rate $R_{ns}$ is calculated in Eq. \eqref{eq: rate_non_sys}. So the difference between these two rates is
\begin{align*}
     & R_{ns} - R_s\\
     =&   \left(1-\frac{\log_2 (2l+1)}{\log_2 k}\left(1-\frac{m}{n}\right)\right)
     - \left( 1-\frac{r}{m+r} \right)\\
      =& \left(1-\frac{\log_2 (2l+1)}{\log_2 k}\frac{gr}{n}\right)
      -\left(1-\frac{r}{n-(g-1)r} \right)\\
      =& \frac{\frac{r}{n}}{1-(g-1)\frac{r}{n}}
      -\frac{\log_2 (2l+1)}{\log_2 k}\frac{gr}{n}\\
      =& \left( \frac{1}{1-(g-1)\frac{r}{n}} - \frac{\log_2 (2l+1)}{\log_2 k}g \right)\frac{r}{n},
\end{align*}
where $l,k$ and $g$ are constant.
When $\frac{r}{n} \to 0$, the difference between the systematic and non-systematic code rates linearly correlates with the changes in $\frac{r}{n}$, with a slope of $ \frac{1}{1-(g-1)\frac{r}{n}} - \frac{\log_2 (2l+1)}{\log_2 k}g  \approx 1 - \frac{\log_2 (2l+1)}{\log_2 k}g$. By Theorem \ref{thm:gray_exist}, the smallest $k$ such that Gray code exists satisfies $\binom{k+3}{3} \le (2l+1)^{3g} E_{2l}$, or $\log_2 k \le (g+1) \log_2 (2l+1)$, where we ignore the  lower order terms for large $l$. When $\log_2 k$ is much larger than $\log_2 l$, this inequality implies that $g$ can be made arbitrarily large. Therefore, the slope can be approximated as $1 -\frac{\log_2 (2l+1)}{\log_2 k}g \le 1 - \frac{g}{g+1} \to 0$, where the limit is taken when $g+1 \ge \frac{\log_2 k}{\log_2 l} \to \infty$. 
This indicates that under conditions of high code rate and large resolution, the code rate is not impaired when adopting the systematic coding. 

\section{Conclusion}\label{sec:conclusion}
This paper proposes a new channel model motivated by composite DNA-based storage. Different from traditional channels, the symbols are probability vectors and the channel noise is modeled as limited-magnitude probability errors. We propose a two-layer error correction code framework that involves classifying and encoding symbols. 
A notable feature of our classification method is its utilization of the characteristics of limited-magnitude probability errors, which helps reduce redundancy and computational complexity. One of the code constructions also exhibits asymptotic optimality. To enhance the practicality of the error correction codes, we also present a systematic code construction. Furthermore, exploring alternative classification methods and designing specialized codes for specific probability error patterns are intriguing avenues for future research. 
The proposed models in this paper have the potential to be extended and applied to other problem domains where information can be represented by probability distributions, thereby opening up new possibilities for its application. 

\appendices

\section{Sphere-packing bound}\label{A}
Here, we provide the derivation of the sphere-packing upper bound for $A_{prob, LMPE}(n,k,t,l)$ in Theorem \ref{thm:sphere_packing}. Our discussion focuses on the case with large resolution $k$. Our approach involves utilizing the total alphabet size, denoted as $|\mathcal{X}^n|$, and the lower bound of error ball to bound the maximum number of possible codewords.

Let $E_{\min}$ be the smallest number of possible errors of maximum magnitude $l$ in a probability vector, then the error ball centered at a given word of length $n$ is of size
\begin{align}
    |\text{error ball}| \ge \sum_{t'=0}^{t} \binom{n}{t'} E_{\min} ^{t'}. \label{eq:49}
\end{align}

\begin{lemma}
    Assume $k \ge 4l$. The smallest number of possible $l$-limited-magnitude probability errors for a probability vector is obtained when the probability vector is $(0,0,0,k)$, and the associated number of errors is
    \begin{align*}
        E_{\min} = \frac{1}{6}l^3+l^2+\frac{11}{6}l+1.
    \end{align*}
\end{lemma}
\begin{IEEEproof}
Denote by $\mathbf{e}=(e_1,e_2,e_3,e_4)$ a valid error vector for $(0,0,0,k)$, which satisfies
   \begin{align}
        &\sum_{i=1}^{4} e_{i}=0,\label{eq: e1}\\
        &0 \leq e_1, e_2, e_3 \leq l,  -l \leq  e_4 \leq 0. \label{eq: range_e}
          \end{align} 
We will show that $(0,0,0,k)$ has the minimum number of valid errors. In particular, we will show that
$\mathbf{e}$ satisfying \eqref{eq: e1}, \eqref{eq: range_e} is also a valid error for an arbitrary probability vector $\mathbf{x}=(x_1,x_2,x_3,x_4)$, when $k\geq 4l$.    
Namely, we will prove that the vector $\mathbf{x}'= \mathbf{x}+\mathbf{e}$ satisfies
\begin{align}
       &\sum_{i=1}^{4} x_{i}'=k, \label{eq: validpro1}\\
       &0 \leq x_i' \leq k, \forall i \in [4]. \label{eq: validpro2} 
\end{align}

Without loss of generality, we set $x_1 \leq x_2 \leq x_3 \leq x_4$. Then we have
   \begin{align}
    \sum_i^{4} x_i =k \label{eq: x},
    \end{align}
where the range of $x_i$ is the following:
\begin{align}
       0\leq x_1\leq \frac{k}{4},\label{eq: x_1}\\
       0\leq x_2\leq \frac{k}{3},\label{eq: x_2}\\
       0\leq x_3\leq \frac{k}{2},\label{eq: x_3}\\
      \frac{k}{4} \leq x_4\leq k .\label{eq: x_4}
\end{align}

Because of \eqref{eq: e1} and \eqref{eq: x}. it is obvious that \eqref{eq: validpro1} is satisfied. Moreover, based on \eqref{eq: range_e}, \eqref{eq: x_1}-- \eqref{eq: x_4}, and $k \ge 4l$, we can prove \eqref{eq: validpro2}:
\begin{align}
           &0 \leq x_1'= x_1+e_1\leq  \frac{k}{4}+l\leq k  ,\label{eq: x_1'}\\
           &0 \leq x_2'= x_2+e_2\leq \frac{k}{3}+l\leq k , \label{eq: x_2'}\\
           &0 \leq x_3'= x_3+e_3\leq \frac{k}{2}+l\leq k  , \label{eq: x_3'}\\
           &0 \leq \frac{k}{4}-l\leq x_4'= x_4+e_4 \leq k \label{eq: x_4'}.
\end{align}

The number of errors associated with $(0,0,0,k)$ is 
\begin{align}
    E_{\min}&=  \sum_{i=0}^{l} \binom{i+2}{2}\\
    &=   \frac{1}{6}l^3+l^2+\frac{11}{6}l+1.\label{eq:E2}
\end{align}
Here, $i$ denotes the sum of the upward or downward error magnitude. 
\end{IEEEproof}

When $k \ge 4l$, the sphere-packing bound for $A_{prob, LMPE}(n,k,t,l) $  implies
\begin{align}
&A_{prob, LMPE}(n,k,t,l)  \\
\leq& \frac{|\mathcal{X}^n|}{|\text{error ball}|}\\
\leq& \frac{|\mathcal{X}^n|}{\sum_{t'=0}^{t} \binom{n}{t'} E_{\min} ^{t'}}\label{eq: 66}\\
\le & \frac{\binom{k+3}{3}^{n}}{\sum_{t'=0}^{t} \binom{n}{t'} \left(\frac{1}{6}\right)^{t'} l^{3t'}} \label{eq: 67}\\
\leq & \frac{\binom{k+3}{3}^{n}}{ \binom{n}{t} \left(\frac{1}{6}\right)^{t} l^{3t}}\label{eq: 68}.
\end{align}
Here, \eqref{eq: 66} is due to \eqref{eq:49}, \eqref{eq: 67} follows because $E_{\min} \ge \frac{l^3}{6}$, and \ref{eq: 68} follows since we only keep the last term of $t=t'$ in the denominator. Moreover, when $l$ is small, the above bound can be modified by applying the exact formula of $E_{min}$ in \eqref{eq:E2}.

\section{Gilbert-Varshamov bound}\label{app:Gilbert}

Here, we provide the derivation of the Gilbert-Varshamov bound for 
$A_{prob, LMPE}(n,k,t,l)$ in Theorem \ref{thm:Gilbert}. 
Our approach involves utilizing the total alphabet size, denoted as $\mathcal{X}^n$, and the upper bound of error ball to bound the minimal number of possible codewords to correct $(l,t)$ LMPE. 

Let $l'$ be the maximum error magnitude for a symbol. Then there can be 0, 1, 2, or 3 positive upward errors (correspondingly 0, 3, 2, or 1 non-negative downward errors, respectively), and the number of possible errors is upper bounded as 

\begin{align}
     E &\le  \sum_{j=1}^{l'} \left(\binom{4}{1} \binom{j+2}{2}+ \binom{4}{2} \binom{j-1}{1} \binom{j+1}{1}+ \binom{4}{3} \binom{j-1}{2} \right) + 1 
    \\
    &= \frac{10}{3}l'^3+5l'^2+\frac{11}{3}l' + 1.\label{eq:E1}
\end{align}
Here, $j$ denotes the sum of the upward or downward error magnitude. When $l' \gg 1$, we approximate the possible error patterns as 
\begin{align}
E  \lesssim \frac{10}{3}l'^3,\label{eq:E_approx}
\end{align}
here, we write $A \lesssim B$ if $\lim_{l \to \infty}\frac{A}{B} \leq 1$.

Due to Theorem \ref{thm:distance}, the geodesic distance being at least $2t+1$ is sufficient to correct $(l,t)$ LMPE. Hence, the Gilbert-Varshamov bound implies
\begin{align*}
A_{prob, LMPE}(n,k,t,l)  &\ge \frac{|\mathcal{X}^n|}{|\text{ball of radius } 2t|}.
\end{align*}
In the denominator, the ball of radius $d=2t$ whose center is a codeword (in geodesic distance for integer $l$) can be partitioned into $d$ different error cases. For $1 \le i \le d$, the $i$-th case is that $i$ symbols have errors whose sum magnitude is at most $dl$, and each symbol's error magnitude is at most a positive multiple of $l$. 
By \eqref{eq:E_approx}, the volume of the set of errors in the $i$-th case can be represented as:
\begin{align*}
  |\text{errors in the $i$-th case}| &\lesssim  \binom{n}{i} \sum_{(l_1,\dots,l_i)}  \left(\frac{10}{3}\right)^i l_1^3...l_i^3 \triangleq  V(i),
\end{align*} 
where the parameter $l_j$ is the maximum magnitude of the $j$-th error, for $j \in [i]$. The summation is over all vectors $(l_1,l_2,\dots,l_i)$ such that each $l_j$ is a positive multiple of $l$, and $\sum_{j=1}^i l_j = dl$. The number of such vectors is $\binom{d-1}{i-1}$. 

Next, we show that $V(i)$ is an increasing function for large $n$. We first find a lower and an upper bound of $V(i)$.

Based on the inequality of arithmetic and geometric means, we have the following inequality:
\begin{align*}
    l_1l_2 \dots l_i \leq \left(\frac{l_1+l_2+\dots+l_i}{i}\right)^{i}=\left(\frac{dl}{i}\right)^{i} ,
\end{align*} 
with equality if and only if $l_1=l_2=\dots=l_i$. Then the upper bound of $V(i)$, denoted by $U(i)$, can be represented as:
\begin{align*}
    U(i) =  \binom{n}{i} \binom{d-1}{i-1} \left(\frac{10}{3}\right)^i\left(\frac{dl}{i}\right)^{3i} .
\end{align*} 
Next, we obtain a lower bound: 
\begin{align*}
    l_1l_2 \dots l_i \geq (d-i+1)l^i,
\end{align*} 
with equality if and only if $l_j=l, j\in[i-1],$ $l_i=(d-(i-1))l$.

So the lower bound of $V(i)$, denoted by $L(i)$, is:
\begin{align}\label{eq:lower bound}
    L(i) = \binom{n}{i}  \binom{d-1}{i-1} \left(\frac{10}{3}\right)^i (d-i+1)^3 l^{3i}.
\end{align} 
Then we consider the following ratio 
\begin{align}
    \frac{V(i+1)}{V(i)} \le \frac{L(i+1)}{U(i)} =  \frac{10}{3} \frac{n-i}{i(i+1)}(d-i)^4 l^3  \left(\frac{i}{d}\right)^{3i},\label{eq: ratio0}
\end{align}
where $i \in  [d-1] $. It can be seen that \eqref{eq: ratio0} is larger than $1$, when $n$ is large, more specifically, when
\begin{align}
  n > \frac{3}{10l^3} \frac{i(i+1)}{(d-i)^4} \left(\frac{d}{i}\right)^{3i} + i. \label{eq:82}
\end{align}
Therefore, $V(i)$ increases as $i$ increases, and thus $V(i) < V(d)$, for large $n$. 
In the expression of \eqref{eq:82}, $\frac{i(i+1)}{(d-i)^4} \le d^2-d$ 
with equality when $i=d-1$. And $\left(\frac{d}{i}\right)^{3i} \le e^{(\frac{3d}{e})}=3.02^d$ with equality when $i=\frac{d}{e}$. Thus, we require that $n>\frac{1}{l^3}d^2 3.02^d$. In this case, the volume of the ball with radius $d$ is upper bounded as 
\begin{align*}
    |\text{ball of radius } d|
    &\lesssim \sum_{i=1}^{d}V(i) < d V(d) = d \binom{n}{d} \left(\frac{10}{3}\right)^d l^{3d}.
\end{align*}
 
For $n>\frac{1}{l^3}d^2 3.02^d, l \gg 1$, setting $d=2t$, we obtain the Gilbert-Varshamov bound: 
\begin{align*}
&A_{prob, LMPE}(n,k,t,l) \\
&\ge \frac{|\mathcal{X}^n|}{|\text{ball of radius } 2t|} \\
&\gtrsim  \frac{\binom{k+3}{3}^{n}}{(2t) \binom{n}{2t} \left(\frac{10}{3}\right)^{2t} l^{6t}}.
\end{align*}

\section{Improved Hamming codes}\label{C}
We provide the construction of the parity check matrix $H$ for improved Hamming codes over $GF((2l+1)^3)$ as in Construction \ref{cnstr:improved Hamming} when not all error patterns are possible. The method is inspired by \cite{Improved_haming}.

We first describe the error vector $\mathbf{e}=(e_1,e_2,e_3,e_4)$ that may appear in the remainder vectors:
\begin{align}
    & 0 \le e_j \le 2l, j \in [4] \label{eq:error_rem1}\\ 
    & \sum_{j=1}^4 e_j \equiv 0 \mod (2l+1),\label{eq:error_rem2}\\
    & \sum_{j\in [4]: e_j \le l} e_j \le l. \label{eq:error_rem3}
\end{align}
For example, when $l=1$, the set of errors for the remainder vectors is shown in Table \ref{table: error pattern}.

\begin{figure}[h]
    \centering
     \includegraphics[width=0.8\linewidth]{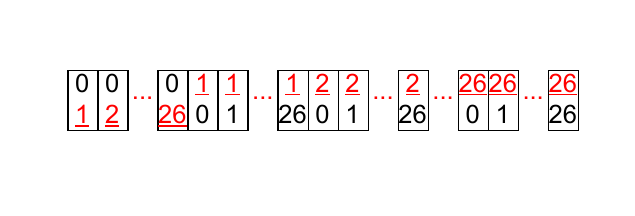}
    \caption{Major elements in all possible columns for $GF(27)$ and 2 parity check symbols. 
    }
    \label{figure: Major elements}
\end{figure}

{\bf Notations.} Consider all columns of length $r$ over $GF(q)$. The \emph{major element} is defined as the first non-zero element in a column, shown as underlined red numbers in Figure \ref{figure: Major elements}. \emph{Major columns} are defined as the columns whose major element is $1$. A \emph{minor column} is defined as a column whose major element is not $1$. Note that the parity check matrix of Hamming code can be constructed by including all major columns.  
Let $\mathcal{E}$ be the subset of $GF(q)$ associated with the possible remainder error patterns.  For example, for $(l=1,t=1)$ LMPE the remainder error patterns are listed in Table \ref{table: error pattern}, and we use the integer representation in the last column to denote $\mathcal{E}$.

\begin{table}
    \centering
    \caption{Remainder error patterns, for $l=1$. The primitive polynomial of $GF(27)$ is chosen to be $x^3+2x+1$. The $3$ coefficients in the polynomial representation correspond to the first $3$ entries of the remainder vector. Every element in $GF(27)$ is represented by an integer between $0$ and $26$, and in particular, each non-zero element is denoted by the integer exponent in the power representation plus $1$.}

    \begin{tabular}{|c| c| c |c|} 
     \hline
     Remainder error pattern&Power&Polynomial & Integer\\
    \hline
    0,0,1,2&1&1&1\\ 
     \hline
    0,1,0,2&$\alpha$&$\alpha$&2 \\
      \hline
    1,0,0,2&$\alpha^2$&$\alpha^2$&3  \\
      \hline
    0,1,2,0&$\alpha^{3}$&$\alpha+2$&4\\
     \hline
    1,2,0,0&$\alpha^4$ &$\alpha^2+2\alpha$&5 \\ 
     \hline
    1,0,2,0&$\alpha^{12}$ & $\alpha^2+2$&13 \\ 
     \hline
     0,0,2,1&$\alpha^{13}$&$2$&14\\
      \hline
     0,2,0,1&$\alpha^{14}$&$2\alpha$&15\\
     \hline
     2,0,0,1&$\alpha^{15}$&$2\alpha^2$&16\\
      \hline
    0,2,1,0&$\alpha^{16}$&$2\alpha+1$&17 \\
     \hline
    2,1,0,0&$\alpha^{17}$& $2\alpha^2+\alpha$&18 \\
     \hline
    2,0,1,0&$\alpha^{25}$& $2\alpha^2+1$&26\\
     \hline
\end{tabular}
    \label{table: error pattern}
\end{table}  
The following construction is the procedure to find the parity check matrix of the improved Hamming code based on \cite{Improved_haming}. The idea is to add columns to the parity check matrix of Hamming code, so that the redundancy $r$ is kept but the codeword length $n$ is enlarged. For the sake of completeness, we prove its correctness in Theorem \ref{prop:improved_hamming}. 

\begin{construction}[Parity check matrix of the improved Hamming code]\label{cnst:pcih}
Let $q \leq (2l+1)^3$ and $r$ be fixed integers.
Initialize the parity check matrix $H$ by including all major columns over $GF(q)$ of length $r$. Initialize the set $\mathcal{I}=\{1\}$, where $\mathcal{I}$ represents the set of major elements in $H$.
For each $i \in GF(q)\backslash \{0,1\}$, append to $H$ minor columns with major element $i$ if the following is satisfied: for all $e_1, e_2 \in \mathcal{E}$ and $j \in \mathcal{I}$,
\begin{align} \label{eq:major condition}
    i e_1 \neq j e_2.
\end{align}
Add $i$ to the set $\mathcal{I}$.
\end{construction}
 
\begin{theorem}\label{prop:improved_hamming}
Construction \ref{cnst:pcih} gives a code over $GF(q)$ with $r$ redundant symbols that can correct a single error in $\mathcal{E}$.
\end{theorem}

\begin{IEEEproof}
Let $H$ in Construction \ref{cnst:pcih} be of size $r \times n$.
Consider an error word of length $n$ that is either all zeros or contains a single non-zero element from $\mathcal{E}$. 
Note that the syndrome equals the product of $H$ and the error word.
We will show that the syndromes of such error words are distinct. Then the decoder can correct the error from the syndrome.

It is apparent that only when no error occurs, the syndrome is all zeros. So we consider the syndrome of two distinct single-error vectors, which can be written as $e_1\mathbf{h}'_1, e_2\mathbf{h}'_2$. Here, $e_1,e_2$ are elements of $\mathcal{E}$ representing the error values, and $\mathbf{h}'_1,\mathbf{h}'_2$ are columns of the parity check matrix $H$ representing the error locations. 
Noticing that each column of $H$ can be viewed as a major column multiplied by some non-zero scalar from $GF(q)$, we can rewrite $\mathbf{h}'_1 = i_1 \mathbf{h}_1, \mathbf{h}'_2 = i_2 \mathbf{h}_2$, where $\mathbf{h}_1,\mathbf{h}_2$ are major columns, and $i_1,i_2 \in \mathcal{I}$ are the scalar multipliers. From \eqref{eq:major condition}, we know
\begin{align}\label{eq:major condition 2}
    i_1 e_1 \neq i_2 e_2.
\end{align}

When $\mathbf{h}_1 \neq \mathbf{h}_2$, since any two major columns in $H$ are not linearly dependent, we know $i_1 e_1 \mathbf{h}_1 \neq i_2 e_2 \mathbf{h}_2$, or the syndromes are different.

When $\mathbf{h}_1 = \mathbf{h}_2$, due to \eqref{eq:major condition 2}, we see $i_1 e_1 \mathbf{h}_1 \neq i_2 e_2 \mathbf{h}_2$.
\end{IEEEproof}

As an example, we apply Construction \ref{cnst:pcih} to remainder classes with $(l=1,t=1)$ LMPE and $q=(2l+1)^3=27, r=2$.
In the initial step, we include the $28$ major columns in the parity check matrix:
\begin{align*}
    \begin{bmatrix}
     0&1&1&...&1 \\
     1&0&1&...&26
    \end{bmatrix}.
 \end{align*} 
By Table \ref{table: error pattern} 
and the condition in \eqref{eq:major condition}, minor columns whose major elements are $i=7$ are added to $H$:
\begin{align*}
    \begin{bmatrix}
     0&7&7&...&7 \\
     7&0&1&...&26
    \end{bmatrix}.
 \end{align*}
No additional values of $i$ satisfy the condition in \eqref{eq:major condition}. 

Therefore, the parity check matrix for $r=2$ has the codeword length $n=56$, which can be reorganized in the systematic form below:
\begin{align*}
    \begin{bmatrix}
     0&1&...&1&7&...&7&1&0\\ 
     7&1&...&26&0&...&26&0&1
    \end{bmatrix}.
 \end{align*} 
 
Note that the set $\mathcal{I}$ in the Construction \ref{cnst:pcih} only depends on the errors $\mathcal{E}$, and does not depend on the redundancy $r$. For general redundancy $r$ and $(l=1,t=1)$ LMPE, we can similarly obtain the codeword length $n= 2\frac{q^r-1}{q-1} = \frac{1}{13}\left(27^r-1\right)$, and the code rate
    $1 -\frac{r(q-1)}{2(q^r-1)}$.

In Construction \ref{cnst:pcih}, there can be multiple ways to map remainder error patterns to $GF(q)$, and the resulting improved Hamming code may have different lengths and rates. The following theorem provides the optimal codes.

\begin{theorem}\label{thm:improved_Hamming_rate}
    Fix the error magnitude $l\geq 1$, the field size $q \leq (2l+1)^3$, and the redundancy $r$.
    Overall possible mappings between error vectors and $GF(q)$ in Construction \ref{cnst:pcih},  the optimal code length of the improved Hamming code is 
    \begin{align*}
        n= I_{\max}\frac{q^r-1}{q-1},
    \end{align*}
    for $I_{\max} = \left\lfloor \frac{q-1}{\frac{10}{3}l^3+5l^2+\frac{11}{3}l}\right\rfloor$. In particular, if $q = (2l+1)^3$ is a prime power, then $I_{\max}=2$.
\end{theorem}

\begin{IEEEproof}
    We will prove that the largest possible set $\mathcal{I}$ contains exactly $I_{\max}$ elements and the theorem follows.
    It can be easily seen that the number of remainder vector errors satisfying \eqref{eq:error_rem1}--\eqref{eq:error_rem3} equals the maximum possible number of probability vector errors (see \eqref{eq:E1}).
    Let 
    \begin{align*}
    I_{\max} = \left\lfloor \frac{q-1}{|\mathcal{E}|} \right\rfloor = \left\lfloor \frac{q-1}{\frac{10}{3}l^3+5l^2+\frac{11}{3}l}\right\rfloor.    
    \end{align*}
    
    We first show that there exists a mapping between errors and $GF(q)$ such that $|\mathcal{I}|=I_{\max}$ is attained.
    Note that the elements in $GF(q)$ can be written as $0,1,\alpha,\alpha^2,\dots, \alpha^{q-2}$ using the power representation, where $\alpha$ is a primitive element.
    Map the remainder error patterns to $ \mathcal{E}=\{\alpha^{I_{\max} j}$, $0 \le j\le |\mathcal{E}|-1\}$ and
    set $\mathcal{I}=\{1, \alpha, \alpha^2, \dots, \alpha^{I_{\max}-1}\}$. 
    For any distinct $\alpha^{i_1},\alpha^{i_2} \in \mathcal{I}$, 
    and any distinct $\alpha^{I_{\max}j_1}, \alpha^{I_{\max}j_2}\in \mathcal{E}$,
    we must have 
    \begin{align*}
        \alpha^{i_1} \alpha^{I_{\max} j_1} \neq  \alpha^{i_2} \alpha^{I_{\max} j_2} 
    \end{align*}
    where the exponent of both sides does not exceed $I_{\max}|\mathcal{E}| -1 \le q-2$. Hence \eqref{eq:major condition} is satisfied and $|\mathcal{I}|=I_{\max}$ is achieved.

    Next, we show that $|\mathcal{I}|$ cannot exceed $I_{\max}$. Suppose on the contrary that $|\mathcal{I}| \ge I_{\max}+1$. By Construction \ref{cnst:pcih}, the sets $\mathcal{E}_i \triangleq \{i e: e \in \mathcal{E}\}$, $i \in \mathcal{I}$, should all be disjoint and does not contain $0$. Thus,
    \begin{align*}
        \Big|\bigcup\limits_{i \in \mathcal{I}} \mathcal{E}_i\Big|
        =\sum_{i \in \mathcal{I}} |\mathcal{E}_i|
        = |\mathcal{I}| |\mathcal{E}|
        \ge  (I_{\max}+1)|\mathcal{E}|
        > \frac{q-1}{|\mathcal{E}|}|\mathcal{E}|
        =q-1,
    \end{align*}
    which contradicts the fact that there are at most $q-1$ non-zero elements in $GF(q)$.
    
    In particular, if $q=(2l+1)^3$, then we proof that $I_{\max}=2$. First, we define the fraction  $I_{inv}$,
\begin{align}\label{eq: ratio}
I_{inv}=\frac{(2l+1)^3}{|\mathcal{E}|}&=\frac{\frac{10}{3}l^3+5l^2+\frac{11}{3}l}{(2l+1)^3} \\
    &= \frac{1}{2}- \frac{2l^3+3l^2-2l+\frac{3}{2}}{24l^3+36l^2+18l+3} \label{eq: larger0}\\
    &= \frac{1}{3}+ \frac{2l^3+3l^2+5l-1}{24l^3+26l^2+18l+1} \label{eq: smaller0}.
\end{align}
Based on $l \geq 1$,
\begin{align}
    \frac{2l^3+3l^2-2l+\frac{3}{2}}{24l^3+36l^2+18l+3} >0 \label{eq: larger1},\\
    \frac{2l^3+3l^2+5l-1}{24l^3+26l^2+18l+1}>0 \label{eq: smaller1}.
\end{align}
Then combining Equation \eqref{eq: larger0}, \eqref{eq: smaller0}, \eqref{eq: larger1} and \eqref{eq: smaller1}, we can get $\frac{1}{3}<I_{inv}<\frac{1}{2}$. Since $I_{max}= \left\lfloor \frac{1}{I_{inv}}\right \rfloor$, $I_{max} =2$. The proof is completed.
\end{IEEEproof}

\section{Redundancy for the constructions}\label{app:redundancy}
We derive the rate of general $(l,t)$ for Construction \ref{cnstr:remainder classes} in Theorem \ref{thm: non_sys}.  Specifically, for Construction \ref{cnstr:remainder classes} with BCH code, we provide the redundancy for general $(l,t)$ in Theorem \ref{prop: redundancy for general t}. For   Construction \ref{cnstr:reduced classes}, the redundancy is derived in Theorem \ref{prop: redundancy for general t2}.  For $(l=1,t=1)$, we show the redundancy of all proposed constructions in Theorem \ref{prop: redundancy for t=1}. The rate examples are shown in Table \ref{table:comparsion sys} and the redundancy examples are summarized in Table \ref{table:redundancy}. For simplicity, we assume that the required finite field size, e.g., the number of classes, is a prime power.

\begin{definition}
Define the rate $R$ of a code $\mathcal{C}$ of length $n$ over an alphabet of size $q$ as
\begin{align*}
    R=\frac{1}{n}\log_q |\mathcal{C}|,
\end{align*}
where $|\mathcal{C}|$ is the number of codewords in $\mathcal{C}$.
\end{definition}

Denote the set of probability vectors with resolution $k$ by 
\begin{align}\label{eq:def_P}
    \mathcal{P} = 
    \{(x_1,x_2,x_3,x_4): \sum_{i=1}^4 x_i = k\},
\end{align}
where $0 \le x_i \le k, i \in [4]$.
Denote the set of remainder vectors by
\begin{align}\label{eq:def_R}
    \mathcal{R}=\{(x_1,x_2,x_3,x_4): 
    \sum_{i=1}^4 x_i \equiv k \mod (2l+1)\},
\end{align}
where $0 \le x_i\le 2l, i \in [4]$.
Next, we derive the number of possible quotient vectors, which is denoted as $Q$.
Denote
\begin{align}\label{eq:s}
 s=\left\lfloor \frac{k}{2l+1} \right\rfloor.   
\end{align}
Assume the quotient vector sums to $j$, and then the corresponding number of quotient vectors is $\binom{j+3}{3}$. 
There are two cases for the choices of $j$.
\begin{itemize}
    \item {\bf Case I.} If $6l+3+ (k \mod (2l+1)) \le 8l$, then $j$ can be $s-3,s-2,s-1$ or $s$, subject to the constraint that $j \ge 0$.  Correspondingly, the remainder vector sums to $6l+3+ (k \mod (2l+1)), 4l+2+ (k \mod (2l+1)), 2l+1+ (k \mod (2l+1))$ or $(k \mod (2l+1))$. The total number of quotient vectors is
    \begin{align}
        Q = \sum_{j=\max\{s-3,0\}}^{s} \binom{j+3}{3}
        \le \sum_{j=s-3}^{s} \binom{j+3}{3} .
        \label{eq:27}\
    \end{align}
    \item {\bf Case II.} If  $6l+3+ (k \mod (2l+1))>8l$, then $j$ can be $s-2,s-1$ or $s$, subject to the constraint that $j \ge 0$. 
    Correspondingly, the remainder vector sums to $ 4l+2+ (k \mod (2l+1)), 2l+1+ (k \mod (2l+1))$ or $(k \mod (2l+1))$.
    The total number of quotient vectors is
    \begin{align}
        Q = \sum_{j=\max\{s-2,0\}}^{s} \binom{j+3}{3}
        \le \sum_{j=s-2}^{s} \binom{j+3}{3} .
        \label{eq:28}
    \end{align}
\end{itemize}

The number of the quotient vectors for a fixed remainder vector is denoted as $Q'$.
Similar to \eqref{eq:27}\eqref{eq:28}, $Q'$  can take the following values:
\begin{align}
    & Q'\in  
    \left\{\binom{j+3}{3}, s_{\min} \le j \le s\right\},\label{eq:26}
\end{align}
where $s_{\min}$ is defined as $\max\{s-3,0\}$ for Case I, or $\max\{s-2,0\}$ for Case II. 

\begin{theorem}[Rate for Construction  \ref{cnstr:remainder classes}]\label{thm: non_sys}
Assume in Construction \ref{cnstr:remainder classes} the code in over $GF(q)$ in the first layer has information length $m$, codeword length $n$, and parity length $n-m$. The LMPE code rate can be approximated as $\frac{m}{n} + (1-\frac{\log_2 (2l+1)}{\log_2 k})(1-\frac{m}{n})$, when $\frac{k}{l}$ is sufficiently large.
\begin{IEEEproof}
Similar to Figure \ref{fig:example1}, the parity symbols in the codeword contain the parity remainder vectors and the information quotient vectors. The parity remainder vectors are determined by the information remainder vectors. 
As mentioned in Example \ref{example:1}, the number of messages that can be represented by the quotient vector in a parity symbol equals to the smallest possible $Q'$ for a given remainder vector.
One can show that the smallest $Q'$ occurs when the remainder vector sums to $4l+2+ (k \mod (2l+1))$ or $6l+3+ (k \mod (2l+1))$, depending on whether the latter exceeds $8l$. And the corresponding number of possible information quotient vectors is $\binom{s_{\min}+3}{3}$. 
The information in the parities can be represented as  $(n-m)\log_2 \binom{s_{\min}+3}{3}$ and the code rate can be calculated as
\begin{align}\label{eq:110}
     \frac{m\log_2 \binom{k+3}{3} + (n-m)\log_2 \binom{s_{\min}+3}{3}}{n\log_2 \binom{k+3}{3}}.
\end{align}
When $\frac{k}{l}$ is large, the code rate can be approximated as  
\begin{align}\label{eq: rate_non_sys}
    \frac{m}{n} + (1-\frac{\log_2 (2l+1)}{\log_2 k})(1-\frac{m}{n}) .
\end{align}
The proof is completed.
\end{IEEEproof}

\end{theorem}

\begin{theorem}[Redundancy for Construction  \ref{cnstr:remainder classes} with BCH code]\label{prop: redundancy for general t}
The redundancy in bits for Constructions \ref{cnstr:remainder classes} with BCH code approaches $2t\log_2(n+1)$ when $\frac{k}{l}$ is sufficiently large.
\end{theorem}

\begin{IEEEproof}
BCH code has length $q^w-1=n$, for the field size $q=(2l+1)^3$. The number of redundant symbols is no more than $r = 2wt = 2t\log_q(n+1)$. Similar to \eqref{eq:110}, the total number of possible words is $\binom{k+3}{3}^n$, the number of information messages is $\binom{k+3}{3}^{n-r}\binom{s_{\min}+3}{3}^r$, and the number of redundant bits is
\begin{align}
    &\log_2 \binom{k+3}{3}^n - \log_2 \binom{k+3}{3}^{n-r}\binom{s_{\min}+3}{3}^r \label{eq:112}\\
    =& \log_2 \frac{\binom{k+3}{3}^{r}}{\binom{s_{\min}+3}{3}^r}\\
    \approx & r \log_2(2l+1)^3 \label{eq:54}\\
    =& 2t \log_2 (n+1). 
\end{align}
Here, \eqref{eq:54} follows when $\frac{k}{l}$ is large.
\end{IEEEproof}

\begin{theorem}[Redundancy for Construction \ref{cnstr:reduced classes} with BCH code]\label{prop: redundancy for general t2}
The redundancy in bits for  Construction \ref{cnstr:reduced classes} approaches $3t\log_2(n+1)$ when $\frac{k}{l}$ is sufficiently large.
\end{theorem}

\begin{IEEEproof}
Similar to Construction \ref{cnstr:remainder classes} with BCH code, the information not only exists in information symbols but also in parity symbols. 

In Construction \ref{cnstr:reduced classes}, BCH code in the first layer has length $q_1^{w_1}-1=n$, for the field size $q_1=(2l+1)^2$. The number of redundant symbols is no more than $r_1=2w_1t=2t\log_{q_1}(n+1)$. 
BCH code in the second layer has length $q_2^{w_2}-1=n$, field size $q_2=2l+1$, and distance $t+1$. The number of redundant symbols is $r_2=w_2t = r_1$. A symbol contains 3 parts: the class index (coded by BCH code in the first layer), the first entry of the remainder vector (coded by BCH code in the second layer), and the quotient vector (information). Since the redundancy length is the same for both BCH codes, the $r_1$ parity symbols contain parity remainder vectors and information quotient vectors, and the remaining $n-r_1$ symbols contain only information. See Figure \ref{fig:reduced classes} (a).
Thus, the number of information messages is $\binom{k+3}{3}^{n-r_1}\binom{s_{min}+3}{3}^{r_1}$, and the number of redundant bits has the same expression as \eqref{eq:54} except that $r$ is different:
\begin{align}
     & r_1 \log_2(2l+1)^3 \label{eq:55}\\
    =& 3t \log_2 (n+1).
\end{align}
The proof is completed.
\end{IEEEproof}

Below we derive the redundancy for naive Hamming code (applied to the entire symbols), Hamming code (applied to the remainder classes as in Construction \ref{cnstr:remainder classes}), improved Hamming code (Construction \ref{cnstr:improved Hamming}), and Hamming code with reduced classes (Construction \ref{cnstr:reduced classes}), which are shown in Table \ref{table:redundancy}. 

\begin{theorem}\label{prop: redundancy for t=1}
Let $l=1,t=1$. The redundancy in bits for naive Hamming code, Hamming code, improved Hamming code, and Hamming code with reduced classes is $\log_2\left(\left(\binom{k+3}{3}-1\right)n+1\right)$,  $\log_2(26n+1)$, $\log_2(13n+1)$, and $\log_2(8n+1)+\log_2(3)$, respectively, when $\frac{k}{l}$ is sufficiently large.
\end{theorem}

\begin{IEEEproof}
Similar to the proof of Theorem \ref{prop: redundancy for general t}, for the parity symbols, 
the number of possible information quotient vectors is $\binom{s_{\min}+3}{3}$, where $s_{\min}=\left\lfloor\frac{k}{2l+1}\right\rfloor-2$. 

Hamming code over $GF(q)$ has length $n=\frac{q^r-1}{q-1}$ always, and the number of redundant symbols is $r=\log_q((q-1)n+1)$. 

{\bf Redundancy for naive Hamming code.} The naive Hamming code requires a field size $q = \binom{k+3}{3}$. The number of redundant symbols $r$ can be represented as $r = \log_q\left(\left(\binom{k+3}{3}-1\right)n+1\right)$. The total number of possible words is $\binom{k+3}{3}^n$, and the number of information messages is $\binom{k+3}{3}^{n-r}\binom{s_{min}+3}{3}^r$. Similar to \eqref{eq:54}, the number of redundant bits is $\log_2\left(\left(\binom{k+3}{3}-1\right)n+1\right)$. 

{\bf Redundancy for Hamming code.} The remainder classes require a field size $q=(2l+1)^3 = 27$. The number of redundant symbols is $r = \log_q(26n+1)$. The number of information messages is $\binom{k+3}{3}^{n-r}\binom{s_{min}+3}{3}^r$. Similar to \eqref{eq:54},  the number of redundant bits is $\log_2(26n+1)$.

{\bf Redundancy for improved Hamming code.} The improved Hamming code has length $n=2\frac{(q^r-1)}{q-1}$ with $q= (2l+1)^3$. The number of redundant symbols is $r = \log_q(\frac{q-1}{2}n+1)$. The number of information messages is $\binom{k+3}{3}^{n-r}\binom{s_{min}+3}{3}^r$. Similar to \eqref{eq:54}, the number of redundant bits is $\log_2(\frac{q-1}{2}n+1)$. For $q=27$, the number of redundant bits of $(l=1,t=1)$ improved Hamming code is $\log_2(13n+1)$.

\begin{figure}
    \centering  \includegraphics[width=0.8\textwidth]{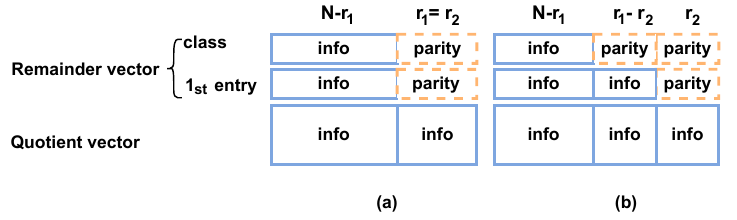}
    \caption{Code structures for Construction \ref{cnstr:reduced classes} with reduced classes. (a) BCH code. (b) Hamming code. }
    \label{fig:reduced classes}
\end{figure}

{\bf Redundancy for Hamming code with reduced classes.} The Hamming code for the first layer has the field size $q_1=9$. The number of redundant symbols is $r_1=\log_{q_1}(8n+1)$. 
The code in the second layer has the field size $q_2=3$, and distance $t+1=2$, which is a single parity check code. The number of redundant symbols is $r_2=\log_{q_2}(3)=1$. 
A symbol contains 3 parts: the class index (9 possibilities) coded by the Hamming code in the first layer, the first entry of the remainder vector (3 possibilities) coded by the single parity check code in the second layer, and the information quotients.  
The first $n-r_1$ symbols are information, the next $r_1-r_2$ symbols contain parity class indices, and the last $r_2$ symbols contain parity remainders, shown in Figure \ref{fig:reduced classes} (b).
The number of possible information messages is $\binom{k+3}{3}^{n-r_1}\binom{s_{min}+3}{3}^{r_1} 3^{r_1-r_2}$.
Similar to \eqref{eq:54}, we can approximate the number of redundant bits for large $\frac{k}{l}$, which is 
\begin{align*}
    & \log_2\binom{k+3}{3}^n-\log_2\binom{k+3}{3}^{n-r_1}\binom{s_{min}+3}{3}^{r_1} 3^{r_1-r_2}\\
    \approx & r_1\log_{q_1} (2l+1)^3 - (r_1-r_2)\log_2 (3)\\
    =& \log_2(8n+1)+\log_2(3).
\end{align*}
The proof is completed.
\end{IEEEproof}

\bibliographystyle{IEEEtran}

\bibliography{bibligraphy.bib}

\end{document}